\documentclass[pra,twocolumn,showpacs,amsmath,amssymb]{revtex4}
\usepackage[latin2]{inputenc}
\usepackage{amsmath}
\usepackage{graphicx}
\usepackage{dcolumn}
\usepackage{epstopdf}

\newcommand{\ba}{\begin{eqnarray*}}
\newcommand{\ea}{\end{eqnarray*}}
\newcommand{\baa}{\begin{eqnarray}}
\newcommand{\eaa}{\end{eqnarray}}
\def\bar{\begin{array}}
\def\ear{\end{array}}
\def\LB{\left(}
\def\RB{\right)}

\def\pr{^{\prime}}

\def\f{\frac}

\begin{document}

\title{Hamiltonian formulation of nonequilibrium quantum dynamics: \\ 
geometric structure of the BBGKY hierarchy
}

\author{Ryan Requist}
\affiliation{
SISSA, via Bonomea 265, 34136 Trieste, Italy 
}

\date{\today}

\begin{abstract}

Time-resolved measurement techniques are opening a window on nonequilibrium quantum phenomena that is radically different from the traditional picture in the frequency domain.  The simulation and interpretation of nonequilibrium dynamics is a conspicuous challenge for theory.  This paper presents an approach to quantum many-body dynamics that is based on a Hamiltonian formulation of the Bogoliubov-Born-Green-Kirkwood-Yvon (BBGKY) hierarchy of equations of motion for reduced density matrices.  These equations have an underlying symplectic structure, and we write them in the form of the classical Hamilton equations for canonically conjugate variables.  Applying canonical perturbation theory or the Krylov-Bogoliubov averaging method to the resulting equations yields a systematic approximation scheme.  The possibility of using memory-dependent functional approximations to close the Hamilton equations at a given level of the hierarchy is discussed.  The geometric structure of the equations gives rise to \textit{reduced} geometric phases that are observable even for noncyclic evolutions of the many-body state.  The approach is applied to a finite Hubbard chain which undergoes a quench in on-site interaction energy $U$.  Canonical perturbation theory, carried out to second order, fully captures the nontrivial real-time dynamics of the model, including resonance phenomena and the coupling of fast and slow variables.

\end{abstract}


\maketitle

\section{Introduction \label{sec:introduction}}

Nonequilibrium quantum dynamics is an important frontier in contemporary physics.  While traditional experimental methods usually probe quantum dynamics in the frequency domain, recent advances in time-resolved measurement techniques have made it possible to study quantum systems on ultrafast time scales.  Noteworthy examples are attosecond pump-probe imaging of electron dynamics \cite{reider2004,corkum2007,krausz2009}, time-of-flight measurements of ultracold atoms in optical lattices \cite{bloch2008}, and ultrafast magnetization dynamics \cite{kirilyuk2010,radu2011}. These experiments and others offer the chance to directly explore little understood topics such as the role of many-body correlation and coherence in real-time dynamics, nonequilibrium quantum quench dynamics \cite{polkovnikov2005,zurek2005,cazalilla2010}, and relaxation in closed quantum systems \cite{biroli2010,polkovnikov2011}. They have also raised the exciting possibility of realizing fundamentally new dynamical phenomena that have no analogs in equilibrium systems. 

These experimental achievements are triggering a renaissance in the theory of nonequilibrium quantum dynamics.  We can now add to the traditional formulations --- nonequilibrium Green function theory \cite{kadanoff1962} with the Keldysh technique \cite{keldysh1964}, the Bogoliubov-Born-Green-Kirkwood-Yvon (BBGKY) hierarchy of equations of motion for reduced density matrices \cite{bogoliubov1946,bogoliubov1961a,bonitz1998}, and time dependent density functional theory (TD DFT) \cite{runge1984} --- a number of sophisticated approximation schemes, including 
nonequilibrium dynamical mean-field theory \cite{freericks2006,eckstein2010},
the time-dependent Gutzwiller approximation \cite{schiro2010,schiro2011}, 
the time-dependent density-matrix renormalization group method \cite{schollwoeck2005}, and 
continuous-time quantum Monte Carlo \cite{gull2011}.  Each of these approximations has strengths and weaknesses, and no practical approach has been found for treating all of the open questions mentioned above.  Keldysh Green function theory is probably the most widely used method for nonequilibrium many-body dynamics.  However, it has drawbacks since in practical calculations one is limited to relatively short times due to the appearance of secular terms \cite{moeckel2010}, i.e.~errors that grow as a power of time, in diagrammatic perturbation theory.  TD DFT is an increasingly popular approach, especially in nanoscale and molecular physics.  In principle, TD DFT and related functional theories would provide a more economical description of real-time quantum dynamics; however, one first needs a functional approximation for the exchange-correlation potential $v_{xc}$ and very little is known about the memory dependence this functional must have in strongly-driven nonadiabatic regimes. 

In this paper, I present a theoretical framework for nonequilibrium quantum dynamics that is based on a Hamiltonian formulation of the BBGKY hierarchy of equations of motion.  The equations are transformed to \textit{classical} Hamilton equations for generalized coordinates and momenta by appealing to the underlying symplectic structure of quantum dynamics.  The principal advantage of writing the equations in this form is the ability to make powerful analogies with the well-developed approximation schemes of classical mechanics.  Applying canonical perturbation theory to the Hamilton equations for reduced variables, we obtain a systematic approximation scheme that goes beyond mean-field theory.  The method is especially useful for fast/slow systems, where there is a separation of time scales.  The Krylov-Bogoliubov averaging method \cite{bogoliubov1961b,kevorkian1996,verhulst2005} can be used to derive effective equations for the slowly varying part, the ``guiding center'', of a dynamical variable by averaging over rapid oscillations.  Averaging reduces the complexity of the equations and gives important insights into the dynamics.  The formulation in terms of classical Hamilton equations may prove useful in analyzing the properties of integrable and nearly-integrable systems as well as the transition to quantum chaos.  Additionally, the use of reduced density matrices has important advantages in strongly-correlated systems because there is no need to rely on a noninteracting reference system.  Effective classical dynamical equations have appeared previously in the context of semiclassical, mean-field or variational approximations \cite{smerzi1997,zobay2000,graefe2006,das2006,itin2007,altland2009,keeling2010,schiro2010,sciolla2011}.  Here, in contrast, Hamilton equations are obtained exactly by means of a transformation to canonically conjugate reduced variables.  

Geometry is gaining recognition as a powerful aid in understanding complex quantum systems.  The quantum geometric tensor \cite{provost1980,berry1989} has been used to analyze quantum phase transitions \cite{carollo2005,campos-venuti2007}, and the effect of geometric phase in nonequilibrium phase transitions is beginning to be addressed \cite{tomka2012}.  Examples of induced gauge potentials \cite{mead1979,provost1980,jackiw1988,berry1989,goldhaber2005,abedi2010} are too numerous to list.  Another way geometric phase manifests itself in real-time dynamics is by modifying the Bohr-Sommerfeld-like interference condition in Stueckelberg oscillations \cite{requist2010,gasparinetti2011}.  This paper shows that a new type of \textit{reduced} geometric phase \cite{requist2011} emerges naturally from the Hamiltonian formulation of the BBGKY equations.  The appearance of these geometric structures and the possibility to exploit them in understanding nonequilibrium dynamics is what distinguishes the present approach from the other approaches mentioned above.

The paper is organized as follows.  The BBGKY hierarchy is reviewed in Sec.~2, and its geometric structure is discussed in Sec.~3.  Section 4 introduces an approximation scheme based on applying canonical perturbation theory to the Hamiltonian formulation of the BBGKY equations.  This method is used to describe the real-time dynamics of an interaction quench in an exactly solvable finite Hubbard chain in Sec.~5.  Conclusions and an outlook on possible directions for further work are given in Sec.~6. 

\section{BBGKY hierarchy of equations of motion}

Consider a closed $N$-body system with a Hamiltonian of the form 
\begin{align}
H = \sum_{i=1}^{N} h_i + \sum_{i<j}^N V_{ij},
\end{align}
where $h$ is a one-body operator and $V$ is an interaction operator.  The density matrix of the system obeys the von Neumann equation ($\hbar=1$)
\begin{align}
i \f{d\rho}{dt} = \left[ H, \rho \right] \label{eq:vonNeumann}.
\end{align}
If the system is in a pure state, then $\rho = \left| \Psi \right> \left< \Psi \right|$ and Eq.~(\ref{eq:vonNeumann}) is equivalent to the Schr\"odinger equation apart from the loss of the overall phase of $\left| \Psi \right>$.  In the general case, the system is described by a mixed state
\begin{equation}
\rho = \sum_i w_i \left| \Psi_i \right> \left< \Psi_i \right|, 
\end{equation}
where $w_i$ are statistical ensemble weights that sum to unity.  Such a description is appropriate when the state of the system is incompletely specified.  The $n$-body reduced density matrix ($n$-matrix) is defined by taking the partial trace of $\rho$,
\begin{align}
\rho_n = \binom{N}{n} \mathrm{Tr}_{n+1\ldots N} \rho, \label{eq:rho-n}
\end{align}
where I have adopted the L\"owdin normalization \cite{loewdin1955}.  One of the nice properties of this convention is that $\mathrm{Tr} \rho_1 = N$ is the number of particles, $\mathrm{Tr} \rho_2 = N(N-1)/2$ is the number of pairs, etc.  The particle density is simply $n(x) = \rho_1(x,x) = \langle x | \rho_1 | x \rangle$.  It is natural to interpret the eigenvalues $n_i$ of $\rho_1$ as the mean occupation numbers of single-particle orbitals, $\left|\phi_i\right>$, the eigenfunctions of $\rho_1$.  These single-particle orbitals are called \textit{natural orbitals} \cite{loewdin1955}.  For fermions, the Pauli principle constrains the $n_i$ to lie in the interval $[0,1]$.  Reduced density matrices encapsulate the information about the averages of all possible physical observables (acting locally in time) in an efficient way.  For example, to evaluate the expectation value of any $n$-body observable $A$ it is enough to know $\rho_n$ because
\begin{align}
\langle A \rangle = \mathrm{Tr}\LB A \rho_n\RB \label{eq:exp-val}.
\end{align}
Since in practice we are mainly interested in one- and two-body observables, we need only calculate $\rho_1$ and $\rho_2$. 

The BBGKY hierarchy is a set of coupled equations of motion for the reduced density matrices.  The equation of motion for level $n$ can be derived by taking the partial trace of the von Neumann equation \cite{bonitz1998}.  One obtains the equation of motion 
\begin{align}
i\f{d\rho_n}{dt} &= \sum_{i=1}^n \left[ h_i, \rho_n \right] + \sum_{i>j=1}^n \left[ V_{ij}, \rho_n \right] \nonumber \\
&\quad + (n+1) \sum_{i=1}^n \mathrm{Tr}_{n+1} \left[ V_{i,n+1}, \rho_{n+1} \right]. \label{eq:BBGKY-leveln}
\end{align}
It is not possible to propagate this equation in time without first knowing $\rho_{n+1}$ since it appears on the right-hand side.  This feature is present at every order (except the last), coupling the entire hierarchy into a sequence.  In practical calculations, the hierarchy is usually closed at some order $n$ by expressing $\rho_{n+1}$ in terms of $\rho_k$ of order $k\leq n$.  In the position representation, the first equation of the hierarchy is
\begin{align}
i\partial_t \rho_1(1,1\pr) &= \left[ h(1) - h(1\pr) \right] \rho_1(1,1\pr) \nonumber \\ &\quad + 2 \int d2 \left[ V(1,2) - V(1\pr,2) \right] \rho_2(1,2,1\pr,2), \label{eq:BBGKY-level1}
\end{align}
where $1=(r_1,\sigma_1)$ is a composite position-spin variable and we have suppressed the time dependence of the density matrices and possible time dependence of $h$ and $V$.

\section{Geometric structure of the BBGKY hierarchy}

In this section, after reviewing the symplectic structure of quantum mechanics (Sec.~\ref{ssec:BBGKY:QM}), I show that the BBGKY hierarchy also has an underlying symplectic structure (Sec.~\ref{ssec:BBGKY:symplectic}), which gives rise to a new type of geometric phase (Sec.~\ref{ssec:BBGKY:geometric-phase}).  By symplectic structure we mean the skew-symmetric structure of a manifold that, in physics, is most familiar from Hamiltonian dynamics in phase space.  Recall that the Hamilton equations for an $n$-freedom classical system with canonically conjugate coordinates and momenta $\{q^{\mu},p_{\mu}\}$ can be written as \cite{jose1998}
\begin{equation}
\f{d\xi^{\mu}}{dt} = \omega^{\mu\nu} \f{\partial H}{\partial \xi^{\nu}}, \label{eq:classical1}
\end{equation}
where $\xi^{\mu} = (q_1 \ldots q_n, p_1 \ldots p_n)$, $H$ is the Hamiltonian and $(\omega^{\mu\nu})$ is the skew-symmetric matrix
\begin{equation}
(\omega^{\mu\nu}) = \LB \bar{rr} 
0_n    &    \; I_n  \\ 
-I_n   &    \; 0_n    \ear \RB \label{eq:omega-matrix}
\end{equation}
with $0_n$ and $I_n$ representing the $n$-dimensional null and identity matrices, respectively.  Throughout the paper, I adopt the Einstein summation convention for repeated Greek indices.  

\subsection{\label{ssec:BBGKY:QM} Symplectic structure of quantum mechanics}

As a prelude to the BBGKY hierarchy, we review the symplectic structure of quantum mechanics \cite{dirac1927,kibble1979,rowe1980,heslot1985} following the presentation in Ref.~\onlinecite{rowe1980}.   

The Dirac-Frenkel stationary action principle
\begin{equation}
\delta \int_{t_1}^{t_2} \left< \Psi \right| i \partial_t - H \left| \Psi \right> dt = 0
\end{equation}
subject to $\left|\delta \Psi(t_1)\right> = \left|\delta \Psi(t_2)\right> = 0$ leads to the differential equation (in a coordinate free representation)
\begin{equation}
i \langle d\Psi | \dot{\Psi} \rangle - i \langle \dot{\Psi} | d\Psi \rangle = dE,
\end{equation}
where the dot represents $d/dt$ and $E = \langle \Psi |H|\Psi\rangle$.  Introducing a complete set of local coordinates $(x^1,x^2,\ldots)$ for projective Hilbert space, this equation becomes 
\begin{equation}
-2\, \mathrm{Im} \langle \partial_{\mu} \Psi | \dot{\Psi} \rangle = -2\, \mathrm{Im} \langle \partial_{\mu} \Psi | \partial_{\nu}\Psi \rangle \dot{x}^{\nu} = \partial_{\mu} E, \label{eq:classical2}
\end{equation}
where $\partial_{\mu} = \partial/\partial x^{\mu}$.  Now introduce the symplectic metric $\sigma$ with components
\begin{equation}
\sigma_{\mu\nu} = -2 \,\mathrm{Im} \langle \partial_{\mu} \Psi | \partial_{\nu} \Psi \rangle.
\end{equation}
If the metric is nondegenerate, then the matrix $(\sigma_{\mu\nu})$ is invertible and according to the Darboux theorem, there exists a canonical transformation from the $x^{\mu}$ to canonical coordinates $\xi^{\mu} = (q_1, \ldots, q_n, p_1, \ldots, p_n)$ under which $(\sigma^{\mu\nu})$ transforms to $(\omega^{\mu\nu})$ and Eq.~(\ref{eq:classical2}) takes exactly the form of the Hamilton equations in Eq.~(\ref{eq:classical1}).  
Since Hamilton equations imply symplectic structure, this tells us that quantum dynamics has a symplectic structure.  The origin of symplectic structure can be traced back to the invariance of the Hermitian inner product on Hilbert space \cite{heslot1985,anandan1991} under unitary transformations.  Note that the imaginary number $i$ has been completely removed from the equations.

Now, consider any two Hermitian observables $F$ and $G$.  Expressing their expectation values for a state $\left| \Psi \right>$ as smooth functions of $x^{\mu}$, the Poisson bracket is defined as
\begin{equation}
\left\{F,G\right\} = \sigma^{\mu\nu} \f{\partial F}{\partial x^{\mu}} \f{\partial G}{\partial x^{\nu}},
\end{equation}
and one can show
\begin{align}
\left\{F,G\right\} = \f{1}{i} \left< \Psi \right| \left[ F,G \right] \left| \Psi \right> .
\end{align}
The equation of motion of $F$ is
\begin{align}
\dot{F} = \f{1}{i} \left< \Psi \right| \left[ F,H \right] \left| \Psi \right> = \left\{ F, H \right\},
\end{align}
which gives the Hamilton equations (\ref{eq:classical1}) for $F=\xi^{\mu}$.

\subsection{\label{ssec:vonNeumann:symplectic} Symplectic structure of the von Neumann equation} 

Before continuing to the BBGKY hierarchy, let us extend the results of the previous section to mixed state dynamics governed by the von Neumann equation.  We shall demonstrate the symplectic structure of the dynamics in systems with finite-dimensional Hilbert spaces \cite{chruscinski1991}. 

Consider an $N$-level quantum system.  The density matrix of the system can be diagonalized by a unitary transformation, i.e., there exists $V\in SU(N)$ such that $D = V^{\dag} \rho V$ is a diagonal matrix.  Let $w_1 > w_2 > \cdots w_m \geq 0$ denote the possibly degenerate eigenvalues of $D$.  Let $k_i$ denote the multiplicity of $w_i$.  We have $k_1 + \cdots + k_m = N$, and the condition $\mathrm{Tr} \rho = 1$ implies $k_1 w_1 + \cdots + k_m w_m = 1$.  Explicitly, the matrix $D$ is
\begin{equation}
D = \LB \bar{ccccc}  
w_1 I_{k_1} &  &  &  \\
 &  \ddots &  &  \\
 &  &  & w_m I_{k_m}  \ear \RB.
\end{equation}
The set $\{w_i, k_i\}$, being conserved by the von Neumann equation, defines a subspace 
\begin{equation}
\mathcal{O}_{\rho} = \{\rho\pr \: | \: \rho\pr = U \rho U^{\dag} \; \mathrm{for} \; U \in SU(N)\}
\end{equation}
of the full space of density matrices.  Thus, the phase space of the system is ``stratified'' into subspaces, and the dynamics takes place entirely within a single subspace.

Through the formulas \cite{chruscinski1991}
\begin{align}
\rho &= N^{-1} I_N + i r \nonumber \\
H &= \mathrm{Tr}H I_N + i h, \label{eq:mapping}
\end{align}
every density matrix and Hamiltonian can be put into one-to-one correspondence with elements $r, h \in \mathfrak{su}(N)$, where $\mathfrak{su}(N)$ denotes the space of anti-Hermitian $N\times N$ matrices.  $\mathfrak{su}(N)$ is the Lie algebra of the Lie group $SU(N)$.  The Lie algebra can be thought of as the space spanned by the generators of infinitesimal rotations. For example, the angular momentum operators $iL_x$, $iL_y$, and $iL_z$ span the Lie algebra $\mathfrak{su}(2)$.  Under the mapping (\ref{eq:mapping}), the von Neumann equation becomes
\begin{equation}
\dot{r} = [h, r]. \label{eq:vonNeumann-Lie}
\end{equation}

We want to show that the subspace $\mathcal{O}_{\rho}$ has a symplectic structure.  To see this it is easier to work with $\mathcal{O}_r$, the subspace of $\mathfrak{su}(N)$ into which $\mathcal{O}_{\rho}$ is carried by the mapping (\ref{eq:mapping}).  If $\mathcal{O}_r$ has a symplectic structure then so does $\mathcal{O}_{\rho}$.  To demonstrate that $\mathcal{O}_r$ has a symplectic structure, we need to identify a skew-symmetric matrix $(\omega^{\mu\nu})$.  First, consider the following bilinear mapping \footnote{The mapping $\omega_s$ is the pull-back of the Kirillov form (see Ref.~\onlinecite{chruscinski1991}).} of two vectors:
\begin{equation}
\omega_s(V_1,V_2) = N \,\mathrm{Re}\, \mathrm{Tr} (s [v_1,v_2]), \label{eq:omega-mapping}
\end{equation}
where $s \in \mathcal{O}_r$ and $v_1,v_2 \in \mathfrak{su}(N)$ are defined by $V_i=[v_i,r]$.  The mapping (\ref{eq:omega-mapping}) is skew-symmetric, i.e. $\omega_s(V_1,V_2) = -\omega_s(V_2,V_1)$.   Since for any $s \in \mathcal{O}_r$, $\omega_s$ maps two vectors of the tangent space at $s$ to a scalar, it is the local mapping corresponding to a global mapping $\omega$ from pairs of vector fields over $\mathcal{O}_r$ to functions on $\mathcal{O}_r$.  The global mapping $\omega$ is a symplectic two-form, which implies that $\mathcal{O}_r$ is a symplectic manifold \cite{chruscinski1991} \footnote{A \textit{symplectic two-form} on a manifold $\mathcal{M}$ is a two-form $\omega$ that is 1) closed; $d\omega=0$ and 2) nondegenerate; $\omega(X,Y)=0$ for all $Y\in T_p\mathcal{M}$ implies $X=0$, where $T_p\mathcal{M}$ is the tangent space to the manifold $\mathcal{M}$ at $p\in \mathcal{M}$.  A \textit{symplectic manifold} is a manifold that can be equipped with a symplectic two-form.}.  Therefore, it is possible to find canonical coordinates such that the von Neumann equation takes the form of the Hamilton equations.  The elements of the skew-symmetric matrix $(\omega_{\mu\nu})$ are given by 
\begin{equation}
\omega_{\mu\nu} = \omega_s(\partial_{\mu},\partial_{\nu}), \label{eq:omega-munu}
\end{equation}
where $\partial_{\mu}$ and $\partial_{\nu}$ are vector fields.  To complete the argument, we note that the symplectic two-form $\omega$ induces a Poisson bracket according to the relationship
\begin{equation}
\{f,g\} = \omega(V_g,V_f),
\end{equation}
where $V_f$ and $V_g$ are the vector fields corresponding to smooth functions $f$ and $g$ defined on $\mathcal{O}_r$.  The dynamical equation for any $f$ is $\dot{f} = \{f,h\}$.  The above steps can be generalized to any semi-simple Lie group by replacing the commutator in Eq.~(\ref{eq:vonNeumann-Lie}) with the Lie bracket \cite{chruscinski1991}.

It is instructive to work through an example.  Consider the two-level case, $SU(2)$.  In direct analogy to the usual Bloch sphere construction, the matrices $r,h \in \mathfrak{su}(N)$ can be expressed as
\begin{align}
r &= \vec{r}\cdot i\vec{\sigma} \nonumber \\
h &= \vec{h}\cdot i\vec{\sigma},
\end{align}
where $\vec{\sigma}$ are the Pauli matrices and $|\vec{r}|=|\vec{h}|=1$.  Hence, the von Neumann equation becomes the Bloch equation
\begin{equation}
\dot{\vec{r}} = \vec{h}\times\vec{r}.
\end{equation}
As the Bloch equation conserves $|\vec{r}|$, it is convenient to work with the spherical angle coordinates $(\theta,\varphi)$.  The arguments above guarantee that we can write the dynamical equations in the form of the Hamilton equations: 
\begin{align}
\dot{\varphi} = \omega^{\varphi\theta} \f{\partial \mathcal{H}}{\partial \theta}, \qquad \dot{\theta} = \omega^{\theta\varphi} \f{\partial \mathcal{H}}{\partial \varphi}, \label{eq:Hamilton-spherical}
\end{align}
where $\mathcal{H}= \cos\theta$ is the Hamiltonian function and $(\omega^{\mu\nu})$ is a skew-symmetric matrix whose elements can be calculated directly from Eq.~(\ref{eq:omega-munu}).  An easier way is to read off the elements $(\omega_{\mu\nu})$ from the differential surface form $\omega = \sin\theta\: d\theta \wedge d\varphi$, which gives $\omega_{\theta\varphi} = -\omega_{\varphi\theta} = \sin\theta$.  Then, the identity $\omega^{\alpha\beta} \omega_{\beta\gamma} = \delta^{\alpha}_{\gamma}$ implies $\omega^{\theta\varphi} = -\omega^{\varphi\theta} = -1/\sin\theta$, so we have
\begin{equation}
\left.(\omega^{\mu\nu})\right|_{\theta,\varphi} = \LB \bar{cc} 0 & \f{1}{\sin\theta} \\
-\f{1}{\sin\theta} & 0 \ear \RB.
\end{equation}
But $(\omega^{\mu\nu})$ is not yet in the form of Eq.~(\ref{eq:omega-matrix}) because $(\varphi,\theta)$ are not canonical coordinates.  Since $\mathcal{H}= \cos\theta = Z$ is a constant of the motion while $\varphi$ is an ignorable coordinate, it is easy to see that $(\varphi,Z)$ are canonical coordinates.  Performing a canonical transformation from $(\varphi,\theta)$ to $(\varphi,Z)$, we find that $(\omega^{\mu\nu})$ takes the canonical form in Eq.~(\ref{eq:omega-matrix}) and the Hamilton equations become
\begin{align}
\dot{\varphi} = \f{\partial \mathcal{H}}{\partial Z} = 1, \qquad \dot{Z} = -\f{\partial \mathcal{H}}{\partial \varphi} = 0. \label{eq:Hamilton-canonical}
\end{align}

\subsection{\label{ssec:BBGKY:symplectic} Symplectic structure of the BBGKY hierarchy}

Now let us look for symplectic structure in the BBGKY hierarchy of equations of motion.  If the equations are found to have symplectic structure, the next step will be to ask if they also have Hamiltonian structure.  We will say that they have Hamiltonian structure if they can be written in the form of Hamilton equations.  Marsden, \textit{et al.}~have previously shown that the \textit{classical} BBGKY hierarchy has a Hamiltonian structure with a Lie-Poisson bracket on the dual of the hierarchy Lie algebra \cite{marsden1984}.  They studied the hierarchy as a whole using the theory of momentum maps \cite{guillemin1980}. In addressing the quantum BBGKY hierarchy I will take a different perspective, focusing on the coupling between adjacent levels of the hierarchy and identifying a Hamiltonian structure with an explicit partitioning of the canonical variables.  

The symplectic structure of the BBGKY hierarchy is ultimately a consequence of the symplectic structure of the von Neumann equation from which it is derived.  In simple terms, we can understand the symplectic structure of the von Neumann equation as following from the appearance of the commutator, which induces a Poisson bracket structure on the space of density matrices.  

We begin by looking for a complete set of canonically conjugate variables for the first level of the hierarchy.  Let us assume that $\rho_1=\rho_1(t)$ is known and ask if we can find a Hamiltonian $h=h(t)$ such that the (hypothetical) equation of motion
\begin{equation}
i\f{d\rho_1}{dt} = [h, \rho_1] \label{eq:hypothetical}
\end{equation}
reproduces the dynamics of $\rho_1$.  We immediately see that this cannot be done because this equation incorrectly conserves the eigenvalues of $\rho_1$, i.e., it predicts $\dot{n}_k = 0$.  This means that the equation of motion for $\rho_1$, Eq.~(\ref{eq:BBGKY-level1}), cannot be put in the form of the von Neumann equation for any Hermitian $h$.  Therefore, to demonstrate that the BBGKY hierarchy has symplectic structure, we will have to modify the arguments used in Sec.~\ref{ssec:vonNeumann:symplectic}.  

Equation (\ref{eq:hypothetical}) \textit{can} generate the correct dynamics of all the eigenfunctions $\left|\phi_k\right>$ of $\rho_1$.  The dynamics is described by a unitary time evolution operator $U(t)=U(t,0)$.  Thus, $\rho_1(t)$ is given by
\begin{align}
\rho_1(t) &= \sum_k n_k(t) \left|\phi_k(t)\right>\left<\phi_k(t)\right| \nonumber \\
&= U(t) \Big[ \sum_k n_k(t) \left|\phi_k(0)\right>\left<\phi_k(0)\right| \Big] U^{\dag}(t). \label{eq:rho1-orbital}
\end{align}
This is analogous to the solution of the von Neumann equation, which is also described by a unitary transformation, $\rho(t) = U(t) \rho(0) U^{\dag}(t)$.  In both cases, the motion is generated by the action of a Lie group on its Lie algebra, namely the adjoint representation.  The difference is that in the von Neumann equation the dynamics is confined to a closed and invariant subspace, namely $\mathcal{O}_r$, determined by the set $\{w_i,k_i\}$, while in Eq.~(\ref{eq:rho1-orbital}) the dynamics passes through multiple $\mathcal{O}_r$ subspaces as the eigenvalues change in time \footnote{The degeneracy of the $n_k$ can also change.}.  Nevertheless, locally in time the dynamics in Eq.~(\ref{eq:rho1-orbital}) has the same Lie algebraic structure as the solution of the von Neumann equation, and we conclude that the carrier manifold for the $\left|\phi_k\right>$ dynamics has a symplectic structure.  This means that we can find a complete set of canonically conjugate coordinates $\{q^{\mu},p_{\mu}\}$ describing all of the linearly independent degrees of freedom of the set $\{\left|\phi_k\right>\}$.  

But what about the occupation numbers $n_k$?  What are their conjugate variables?  The variable conjugate to $n_k$ is a phase $\zeta_k$; it a degree of freedom of all $\rho_n$ with $n\ge 2$.   More precisely, $\zeta_k$ is the degree of freedom corresponding to the one-parameter family of unitary transformations $U \rho_n U^{\dag}$, where $U=e^{-i s n_k}$ and $s$ is a parameter.  The operator $n_k$ is the generator of $\zeta_k$ translations in the same way that the momentum operator is the generator of spatial translations.  However, $\zeta_k$ is not the expectation value of any self-adjoint operator because the existence of such an operator would violate the uncertainty principle \cite{carruthers1968}.  The importance of the phases $\zeta_k$ for the dynamics of the $n_k$ was recognized in Refs.~\onlinecite{requist2010}, \onlinecite{requist2011} and \onlinecite{giesbertz2010b}.  

The phases $\zeta_k$ introduced in Refs.~\onlinecite{requist2010} and \onlinecite{requist2011} are not invariant to the gauge transformation $\left|\phi_k\right> \rightarrow e^{i\lambda_k} \left|\phi_k\right>$.  The $\zeta_k$ should be understood as \textit{relative} phases because they can only be uniquely defined relative to a specific choice of time-dependent phases for the $|\phi_k\rangle$.  However, gauge-invariant phases $\alpha_k = \int^t \dot{\alpha}_k dt\pr$ can be defined through the expression 
\begin{equation}
\dot{\alpha}_k = \dot{\zeta}_k + i \langle \phi_k | \dot{\phi}_k \rangle.
\label{eq:alphak}
\end{equation}
The phase $\alpha_k$ is invariant under the above gauge transformation because $i \langle \phi_k | d\phi_k \rangle \rightarrow i \langle \phi_k | d\phi_k \rangle - d\lambda_k$ while $d\zeta_k \rightarrow d\zeta_k + d\lambda_k$.  In Ref.~\onlinecite{giesbertz2010b}, phase-including natural orbitals $e^{-i\zeta_k} |\phi_k\rangle$ (in our notations) are defined, which are gauge invariant for the same reason.  Equation (\ref{eq:alphak}) has the form of a covariant derivative.  In Sec.~\ref{ssec:BBGKY:geometric-phase} we shall show that the $\zeta_k$ combine with the $n_k$ and $\left|\phi_k\right>$ to form a geometric phase.  It is worth noting that any $\alpha_k=\alpha_k(t)$ can be realized if the Hamiltonian, containing one-body and two-body operators, is allowed to be arbitrary.  For any $H=H(t)$ generating phases $\alpha_k(t)$, $H\pr=U H U^{\dag} - i U \partial_t U^{\dag}$ with $U=e^{i\beta_k n_k}$ generates phases $\alpha_k\pr(t) = \alpha_k(t) + \beta_k(t)$.  Notice that such a transformation does not affect the one-body terms of $H$. 

The $\alpha_k$ and $\zeta_k$ do not appear in $\rho_1$.  It is surprising that the degrees of freedom of $\rho_1$ do not form a closed set of canonically conjugate variables.  To construct a complete set of conjugate variables describing all of the degrees of freedom of $\rho_1$, namely the set $\{q^{\mu},p_{\mu},n_k\}$, it is necessary to add the variables $\alpha_k$ to that set.  The $\alpha_k$ can be taken as the conjugate variables of the $n_k$.  Thus, the symplectic structure interweaves the levels of the BBGKY hierarchy.  The $\alpha_k$ correspond to the ``lost'' phase of the natural orbital $\left|\phi_k\right>$; since the $\left|\phi_k\right>$ are defined by the eigenvalue equation $\rho_1 \left|\phi_k\right> = n_k \left|\phi_k\right>$, their phases are undefined.  

To summarize the above paragraphs, the complete set of canonically conjugate variables for the first level of the hierarchy is formed by adding $\{\alpha_k,n_k\}$ to the set $\{q^{\mu},p_{\mu}\}$ representing the eigenstate (orbital) degrees of freedom.  The same structure is repeated at every level of the hierarchy.  For a general level $n$, let $\{\overline{Q}_n^{\mu},\overline{P}_{n\mu}\}$ denote a set of canonically conjugate variables comprising all of the eigenstate degrees of freedom of $\rho_n$.  The complete set of canonically conjugate variables for level $n$ is defined to be $\{Q_n^{\mu},P_{n\mu}\} = \{\overline{Q}_n^{\mu},\overline{P}_{n\mu}\} \cup \{\alpha^{\mu}_n,\lambda_{n\mu}\}$, where $\lambda_{n\mu}$ are the eigenvalues of $\rho_n$ and $\alpha^{\mu}_n$ are their conjugate variables, relative phases of $\rho_{n+1}$.

The hierarchical structure of the BBGKY equations can be used to organize all of the canonically conjugate variables into a hierarchy.  In building such a hierarchical structure, one has to keep in mind that the degrees of freedom of $\rho_n$ are not linearly independent of the degrees of freedom of $\rho_{n-1}$ since the latter can be obtained from the former by the partial trace
\begin{equation}
\rho_{n-1} = \f{n}{N-n+1} \mathrm{Tr}_n \rho_n. \label{eq:partial-trace}
\end{equation}
To handle this interdependency, we can make a canonical transformation of the variables $\{Q_n^{\mu},P_{n\mu}\}$ to a new set of variables $\{Q_{n-1}^{\mu},P_{n-1\mu}\} \cup \{q_n^{\mu},p_{n\mu}\}$, isolating the degrees of freedom of level $n-1$ from the remaining degrees of freedom $\{q_n^{\mu},p_{n\mu}\}$ \footnote{It should be possible to justify this procedure by the second statement of the Darboux theorem (see Ref.~\onlinecite{jose1998}).}.  The set of variables $\{q_{n}^{\mu},p_{n\mu}\}$ is independent of the degrees of freedom of level $n-1$ and, by induction, all lower levels.  Here, the independence of two variables means that their Poisson bracket vanishes.  Starting at the bottom (level 1) and working up, the entire hierarchy can be partitioned into mutually disjoint sets of canonically conjugate variables, each associated with a particular level of the hierarchy:
\begin{equation}
\underbrace{\{q_1^{\mu},p_{1\mu}\} \cup \{q_2^{\mu},p_{2\mu}\}}_{\{Q_2^{\mu},P_{2\mu}\}} \cup \cdots \cup \{q_N^{\mu},p_{N\mu}\} = \{Q_N^{\mu},P_{N\mu}\}. \label{eq:hierarchy:sets}
\end{equation}
Thus, all the degrees of freedom of the density matrix $\rho$ have been organized into a hierarchical set of canonically conjugate variables.  Now we can ask the following three questions.  Is there an effective Hamiltonian function that generates the dynamics of the complete set of variables in Eq.~(\ref{eq:hierarchy:sets})?  What is the form of the coupling between the variables $\{q_{n}^{\mu},p_{n\mu}\}$ and $\{q_{n+1}^{\mu},p_{n+1\mu}\}$?  Does the separation of $\{Q_{n}^{\mu},P_{n\mu}\}$ into $\{Q_{n-1}^{\mu},P_{n-1\mu}\} \cup \{q_{n}^{\mu},p_{n\mu}\}$ induce gauge structure \cite{mead1979,provost1980,jackiw1988,berry1989} in the effective equations of motion for the reduced variables.  The first question will be addressed in sections \ref{sec:CPT} and \ref{sec:eff:Ham}; the second two will be left for future work.  We shall now discuss possible routes to a rigorous proof of the symplectic structure of the BBGKY hierarchy; the uninterested reader may wish to skip ahead to Sec.~\ref{ssec:BBGKY:geometric-phase}.

The arguments presented above for the existence of symplectic structure are not rigorous.  In concluding this section, I want to mention some issues that one might have to confront in formulating a rigorous proof.  The symplectic structure of the von Neumann equation has been established in a quite general case \cite{chruscinski1991}, and it will probably be possible to extend this result to most physically interesting cases by considering infinite-dimensional Lie algebras.  Above, I have claimed that the manifold---let us denote it as $\mathcal{M}_n$---of all eigenfunction degrees of freedom of $\rho_n$ has a symplectic structure.
In order to prove this statement, one has to show that $\mathcal{M}_n$ can be equipped with a symplectic two-form $\omega_n = d\overline{Q}_n^{\mu} \wedge d\overline{P}_{n\mu}$.  Since $\mathcal{M}_n$ is a subspace of the space $\{\rho\}$ of full density matrices, which we know is equipped with a symplectic two-form $\omega$ according to the arguments in Ref.~\onlinecite{chruscinski1991}, the essential question is whether the restriction of $\omega$ to $\mathcal{M}_n$ remains a symplectic two-form.  

To prove that the restricted two-form $\omega_{n}$ is symplectic, one has to show  1) $\omega_{n}$ is closed, i.e.~$d\omega_{n} = 0$ and 2) $\omega_{n}$ is nondegenerate, i.e.~for all $p\in \mathcal{M}_n$ and all $Y\in T_p\mathcal{M}_n$, $\omega_{n}(X,Y)=0$ implies $X=0$, where $T_p\mathcal{M}_n$ is the tangent space to $\mathcal{M}_n$ at $p$.  Condition (1) is almost certainly satisfied due to the linearity of the partial trace in the definition of $\rho_n$.  Condition (2) is more difficult to prove.  If the eigenvalues $\lambda_1$ and $\lambda_2$ corresponding to two eigenvectors $\left|\Phi_1\right>$ and $\left|\Phi_2\right>$ of $\rho_n$ become degenerate at some time, one might expect $\omega_{n}$ to become degenerate, i.e.~one might expect there to exist $X\neq 0$ and $Y \neq 0$ such that $\omega_{n}(X,Y) = 0$.  However, we can easily see that this situation cannot arise if $\mathcal{M}_n$ is properly defined.  We must define $\mathcal{M}_n$ to be the space of all \textit{linearly independent} degrees of freedom associated with the eigenfunctions of $\rho_n$.  For example, if two eigenfunctions are degenerate, then they are only defined up to an $SU(2)$ unitary transformation, i.e.~$\left|\Psi_1\right> = \cos(\theta/2) e^{-i\varphi/2} \left|\Phi_1\right> + \sin(\theta/2) e^{i\varphi/2} \left|\Phi_2\right>$ and $\left|\Psi_2\right> = -\sin(\theta/2) e^{-i\varphi/2} \left|\Phi_1\right> + \cos(\theta/2) e^{i\varphi/2} \left|\Phi_2\right>$ are equally valid eigenfunctions.  The variables $\theta$ and $\phi$ associated with the unitary transformation should not be considered degrees of freedom of the space $\mathcal{M}_n$.  This situation is readily generalized to multiple subsets of degenerate eigenfunctions with any degree of degeneracy.  There is the freedom to make an arbitrary unitary rotation within each degenerate subset.  The important point is that the variables corresponding to these unitary rotations are not degree of freedoms of $\mathcal{M}_n$, so they cannot be a source of degeneracy of $\omega_{n}$.  

In the context of Lie algebras, this situation is easily handled by defining the quotient space $\mathcal{O}_x = G/G_x$, where $G$ is the Lie group and $G_x = \{ U \in G | UxU^{\dag} = x\}$ 
is the isotropy subgroup at an element $x$ of the Lie algebra \cite{chruscinski1991}.  Such a quotient space is called a \textit{flag manifold}.  For the example of Sec.~\ref{ssec:vonNeumann:symplectic}, where $G=SU(N)$, 
\begin{equation}
\mathcal{O}_r = \f{SU(N)}{SU(k_1)\times \cdots \times SU(k_m)}.
\label{eq:flag}
\end{equation}
We must identify $\mathcal{M}_n$ with $\mathcal{O}_r$ not $SU(N)$. 

Another question to ask in connection with the construction of a hierarchy of canonically conjugate variables is whether all of the eigenvalues $\lambda_{nk}$ of $\rho_n$ can be considered as linearly independent degrees of freedom.  It is known that there are certain nontrivial conditions, so-called $N$-representability conditions \cite{coleman1963}, that a given $\rho_n$ must satisfy in order to be obtainable from some $N$-body state $\left| \Psi\right>$.  The general problem of finding explicit constraints guaranteeing that a candidate $\rho_n$ can be obtained from a certain type of $N$-body state is known as the $N$-representability problem or quantum marginal problem.  For $n=1$, the necessary and sufficient conditions for $\rho_1$ to come from an $N$-body ensemble with arbitrary weights $w_i$ are: i) $\rho_1$ is Hermitian, ii) $0\le n_k \le 1$ and iii) $\sum_k n_k = N$ \cite{coleman1963}.  The solution of the $N$-representability problem for a pure state or an ensemble state with given ensemble weights $w_i$ has been reported for $n=1$ \cite{klyachko2004,klyachko2006}; explicit constraints on the $n_k$ are found for given dimension $d$ of the single-particle Hilbert space.  The interesting observation for our purposes is that when $d$ is large enough compared to $N$, all of the constraints take the form of inequalities.  Presumably, as long as the set of occupation numbers $\{n_k\}$ does not lie on the boundary of the $N$-representable region defined by the inequality constraints, the $n_k$ can be considered as linearly independent degrees of freedom.  It is also worth noting that symplectic geometry has very recently been applied to this problem and similar problems \cite{christandl2012}.  At least for the first level of the hierarchy, it appears that we can indeed consider the $n_k$ as linearly independent degrees of freedom.  

If the symplectic two-form $\omega_n = d\overline{Q}_n^{\mu} \wedge d\overline{P}_{n\mu}$ exists, then the two-form 
\begin{equation}
\Omega_n = d\overline{Q}_n^{\mu} \wedge d\overline{P}_{n\mu} + d\alpha^{\mu}_n \wedge d\lambda_{n\mu}
\end{equation}
is a symplectic two-form for the complete space of variables $\{Q_n^{\mu},P_{n\mu}\}$.  The eigenstates $|\Phi_{nk} \rangle$ of $\rho_n$ are elements of a projective Hilbert space, which is a K\"ahler manifold \cite{ashtekar1998,ashtekar1997}.  A K\"ahler manifold is endowed with a Hermitian form $h=g + i \omega$, where $g$ is a Riemannian metric (the Fubini-Study metric) and $\omega$ is a symplectic two-form.  It is an interesting question whether the space $\{Q_n^{\mu},P_{n\mu}\}$ is also a K\"ahler manifold, and if so, what is the physical significance of the metric $g$ and the K\"ahler potential from which it is derived.

\subsection{\label{ssec:BBGKY:geometric-phase} One-body reduced geometric phases}

The phases $\zeta_k$ are not observable because they are not invariant to the gauge transformation $|\phi_k\rangle \rightarrow e^{i \lambda_k} |\phi_k\rangle$, yet globally the functions $\zeta_k(t)$ can be put together with $n_k(t)$ and $|\phi_k(t)\rangle$ to form observable geometric phases.  Consider a cyclic evolution of $(|\phi_k\rangle,n_k,\zeta_k)$ on the time interval $[0,T]$; $|\phi_k(T)\rangle = |\phi_k(0)\rangle$, $n_k(T)=n_k(0)$, and $\zeta_k(T)=\zeta_k(0)$.  We do \textit{not} need to assume that the full density matrix $\rho$ also undergoes a cyclic evolution.  If $n_k$ is nondegenerate, the quantity 
\begin{equation}
\gamma_k = \oint \LB i n_k \langle \phi_k | d\phi_k \rangle + n_k d\zeta_k \RB \label{eq:reduced-geometric-phase}
\end{equation}
is a geometric phase \cite{requist2011}, which we shall refer to as the \textit{one-body reduced geometric phase}.  The first term resembles the expression for the geometric phase \cite{berry1984,aharonov1987} associated with the parallel transport of the orbital $|\phi_k\rangle$ except it is multiplied by $n_k$, which reduces the orbital contribution with respect to its ``bare'' value.  The second term is an extra contribution that depends, through $\zeta_k$, on two-body degrees of freedom.  Due to the presence of the factor $n_k$, neither term is gauge invariant but their sum is \cite{requist2011}.  This is because $\zeta_k$ exactly compensates for the gauge freedom of $|\phi_k\rangle$ (see Sec.~\ref{ssec:BBGKY:symplectic}).  

The physical meaning of $\gamma_k$ can be understood as follows.  First, note that $\rho_1$ contains the information about all one-body observables, such as the density and current density.  If $\rho_1(T)=\rho_1(0)$ at some time $t=T$, then all one-body observables have returned to their initial values.  In such a situation, $\{\gamma_k\}$ are a set of geometric phases that tells us about the path the system took in the space of all possible $\rho_1$.  One can also think of $\gamma_k$ as a geometric phase associated with a cyclic evolution of the single-particle state $|\psi_k\rangle  = e^{-i\zeta_k} \sqrt{n_k} |\phi_k\rangle$ in a projective Hilbert space augmented by a pair of variables which act like a square modulus and phase.

Berry and Aharonov-Anandan phases \cite{berry1984,aharonov1987} require cyclic evolution of the full wave function.  The reduced geometric phases in Eq.~(\ref{eq:reduced-geometric-phase}) only require cyclic evolution of the set of variables $(|\phi_k\rangle,n_k,\zeta_k)$ --- a weaker condition.  Therefore, the reduced geometric phases are observable in situations where the full geometric phase is not.  The reduced geometric phases can be observed in interference experiments.  Consider two final states $|\Psi(T)\rangle$ and $|\Phi(T)\rangle$ differing only in a particular reduced geometric phase $\gamma_k$.  Although both states have exactly the same $\rho_1(T)$, the effect of the reduced geometric phase is observable in the interference of the cross terms of a trial wave function $\alpha |\Psi(T)\rangle + \beta |\Phi(T)\rangle$, where $|\alpha|^2+|\beta|^2=1$.   It would be interesting to study the relationship between the $\gamma_k$ and the Uhlmann geometric phase for mixed states \cite{uhlmann1986,sjoqvist2000a} and the geometric phases of entangled spins \cite{sjoqvist2000b}.  Finally, we mention that it should be possible to extend the definition of the reduced geometric phases to open paths as was done for the full geometric phase \cite{samuel1988}.  

In terms of $|\psi_k\rangle$, Eq.~(\ref{eq:reduced-geometric-phase}) can be expressed in the form of the generalized Stokes theorem
\begin{align}
\gamma_k &= \oint_{\mathcal{C}_k} \beta_k = \int_{\mathcal{S}_k} \omega_k, \label{eq:Stokes}
\end{align}
where $\beta_k = i \langle \psi_k | d\psi_k \rangle$ is a connection one-form, $\omega_k= i \langle d\psi_k | \wedge | d\psi_k \rangle$ is the associated two-form, and $\mathcal{C}_k$ is a closed path bounding the surface $\mathcal{S}_k$ in the space $\{q_1^{\mu},p_{1\mu}\}$.  The geometric phase $\gamma_k$ is a nonintegrable phase that arises due to the nonexactness of the one-form $\beta_k$, i.e., the fact that there does not exist a function $f$ such that $\beta_k = df$.  Owing to the symplectic structure of projective Hilbert space \cite{anandan1991} augmented by the canonically conjugate pair $(\alpha_k,n_k)$, $\gamma_k$ can be expressed as an action integral $\oint p_{\mu} dq^{\mu}$.  The sum $\omega = \sum_k \omega_k$ is the symplectic two-form for the manifold $\{q_1^{\mu},p_{1\mu}\}$, which is similar to a result for mixed states with constant ensemble weights \cite{chaturvedi2004}.   Analogously, $\sum_k \gamma_k$ returns the geometric phase of the full wave function in two-electron systems \cite{requist2011}.  Similar $n$-body reduced geometric phases will arise at higher levels of the BBGKY hierarchy.

\section{\label{sec:CPT} Canonical perturbation theory of the BBGKY hierarchy}

One important benefit of formulating the BBGKY equations as Hamilton equations is the possibility of applying the well-developed classical approximation schemes such as canonical perturbation theory (CPT).  CPT has recently been applied successfully to quantum systems \cite{itin2007,itin2009,itin2011}.  Before introducing the CPT of the BBGKY hierarchy, it is worth briefly mentioning two other approximations that can be applied to the Hamilton equations -- the Krylov-Bogoliubov (KB) averaging method and the separation of fast and slow variables.  In some problems, there are relationships between these three methods.  

The KB averaging method can be used when the solution of the dynamical equations has the form $x = X + \tilde{x}$, where $X$ is smoothly varying and $\tilde{x}$ is the sum of small oscillatory terms.  The name ``averaging'' comes from the fact that, after transforming the equations to the standard form $\dot{x} = \epsilon F(x,t)$, the right-hand side is time averaged to remove all oscillatory contributions to $F$.  In other words, the time average removes all terms except $F_0(x)$ from the Fourier series $F(x,t)=\sum_n e^{i\omega_n t} F_n(x)$.  The oscillatory contributions are accounted for in higher orders.  In the context of the BBGKY equations, averaging might provide a way to derive effective dynamical equations that describe relaxation phenomena.  Since for this purpose one would like to have nonconservative equations, one needs noncanonical transformations.

In methods based on a separation of fast and slow variables, such as the Born-Oppenheimer approximation, one looks for an asymptotic expansion of the equations of motion of two sets of variables whose dynamics take place on different time scales.  In general, one does not know \textit{a priori} which degrees of freedom are fast and which are slow.  The hierarchical structure of the BBGKY equations might help in identifying fast and slow degrees of freedom.  One can envision making either vertical or horizontal separations of the hierarchy.  In a vertical separation, approximations would be based on the fact that the variables of one level (or a subset of such variables) are much faster than those of an adjacent level.  For example, in weakly interacting systems the occupation numbers $n_k$ are weakly driven and hence slowly varying \cite{requist2010}.  In a horizontal separation, a certain subset of variables $\{x_n\} \subset \{q_n^{\mu},p_{n\mu}\}$ of level $n$ would be considered as fast variables.  Depending on the form and strength of the coupling between levels, the $\{x_n\}$ might induce fast motion in a certain subset of variables $\{x_{n+1}\}$ of level $n+1$, and so on up the hierarchy.  

In the present section, we will focus on CPT.  There are many ways that CPT can be applied to the Hamiltonian formulation of the BBGKY hierarchy.  In general, one should look for a solvable zeroth-order Hamiltonian $H_0 = H_0(q_1^{\mu}\ldots q_N^{\mu},p_{1\mu} \ldots p_{N\mu})$ that approximates the dynamics of the full $H$.  Then, the difference $H_1 = H - H_0$ can be treated as a perturbation.  It might be possible to take $H_0$ in the form of Eq.~(\ref{eq:separable}) and treat the coupling between adjacent levels of the hierarchy within CPT.  In this section, we formulate CPT for a general Hamiltonian, assuming for convenience that $H_0$ and $H_1$ are time independent.  This restriction can be removed.  Our presentation will follow Ref.~\onlinecite{jose1998}.

Suppose the Hamiltonian of an $N$-particle system can be written as 
\begin{align}
H &= H_0(q_1^{\mu}\ldots q_N^{\mu},p_{1\mu} \ldots p_{N\mu}) \nonumber \\
&+ \epsilon \, H_1(q_1^{\mu}\ldots q_N^{\mu},p_{1\mu} \ldots p_{N\mu}),
\end{align}
where the dynamics are integrable for $H_0$ and $\epsilon$ is a small parameter introduced for bookkeeping purposes.  Since the dynamics are integrable for $H_0$, there exists action-angle (AA) variables $\{\psi^{\mu}_0,J_{\mu 0}\}$ such that
\begin{align}
\dot{\psi}^{\mu}_0 &= \f{\partial H_0}{\partial J_{\mu 0}} = \omega^{\mu}_0 \qquad \dot{J}_{\mu 0} = \f{\partial H_0}{\partial \psi^{\mu}_0} =0, \label{eq:AA-zeroth-order}
\end{align}
where $\omega^{\mu}_0$ are constant frequencies.  The solution of the original problem is obtained by transforming back to the variables $\{q_1^{\mu}\ldots q_N^{\mu},p_{1\mu}\ldots p_{N\mu}\}$.  In terms of the AA variables, the zeroth-order Hamiltonian is simply
\begin{equation}
H_0(J_{\mu 0}) = \omega^{\mu}_0 J_{\mu 0}.
\end{equation}
Now we want to derive a perturbation series for the dynamics of $H$.  To do so, we start by assuming that the exact dynamics is integrable.  Even if this is not true, the $\epsilon$-series generated in CPT may still prove useful as an asymptotic series.  The assumption that $H$ is integrable implies the existence of AA variables $(\psi^{\mu},J_{\mu})$ that solve the full problem.  To proceed, we write the following power series for the type 2 generating function $S$ of the canonical transformation $(\psi^{\mu}_0,J_{\mu 0}) \rightarrow (\psi^{\mu},J_{\mu})$:
\begin{equation*}
S(\psi^{\mu}_0,J_{\mu}) = \psi^{\mu}_0 J_{\mu} + \epsilon S_1(\psi^{\mu}_0,J_{\mu}) + \epsilon^2 S_2(\psi^{\mu}_0,J_{\mu}) + \cdots . 
\end{equation*}
Once the generating function is known, the series for the AA variables are given by
\begin{align}
\psi^{\mu} &= \f{\partial S}{\partial J_{\mu}} = \psi^{\mu}_0 + \epsilon \f{\partial S_1}{\partial J_{\mu}} + \epsilon^2 \f{\partial S_2}{\partial J_{\mu}} + \cdots \nonumber \\
J_{\mu 0} &= \f{\partial S}{\partial \psi^{\mu}_0} = J_{\mu} + \epsilon \f{\partial S_1}{\partial \psi^{\mu}_0} + \epsilon^2 \f{\partial S_2}{\partial \psi^{\mu}_0} + \cdots, \label{eq:AA-series}
\end{align}
where the right-hand sides are evaluated for $\psi^{\mu}_0$ and $J_{\mu}$.  To obtain $J_{\mu}$ in terms of $J_{\mu 0}$, the power series in the second equation has to be inverted.  The next step is to write a power series for the Hamiltonian.  The integrability of $H$ implies the existence of a Hamiltonian function $E$ that depends only on the $J_{\mu}$.  Let us expand it as follows:
\begin{equation}
E(J_{\mu}) = E_0(J_{\mu}) + \epsilon E_1(J_{\mu}) + \epsilon^2 E_2(J_{\mu}) + \cdots.
\end{equation}
By comparing like powers of $\epsilon$ between this series and a similar power series for the Hamiltonian $H=H_0+\epsilon H_1$, expressed in terms of $\{\psi^{\mu}_0,J_{\mu}\}$, one obtains 
\begin{align}
E_0(J_{\mu}) &= H_0(J_{\mu}) \nonumber \\
E_1(J_{\mu}) &= H_1(\psi^{\mu}_0,J_{\mu}) + \f{\partial H_0}{\partial J_{\mu}} \f{\partial S_1}{\partial \psi^{\mu}_0} \nonumber \\  
E_2(J_{\mu}) &= \f{\partial H_1}{\partial J_{\mu}} \f{\partial S_1}{\partial \psi^{\mu}_0} + \f{\partial H_0}{\partial J_{\mu}} \f{\partial S_2}{\partial \psi^{\mu}_0} + \f{1}{2} \f{\partial^2 H_0}{\partial J_{\mu} \partial J_{\nu}} \f{\partial S_1}{\partial \psi^{\mu}_0} \f{\partial S_1}{\partial \psi^{\nu}_0} \label{eq:E-S}
\end{align}
These expressions give the relationship between the $E_n$ and $S_n$, but so far these are both unknown functions.  This problem can be solved by performing an averaging over the $\psi^{\mu}_0$ dependence.  We define the $\psi_0$-average of a function $F(\psi^{\mu}_0,J_{\mu})$ as
\begin{equation}
\langle F \rangle = \f{1}{(2\pi)^\mathcal{N}} \int_0^{2\pi} \!\cdots \int_0^{2\pi} F(\psi^{\mu}_0,J_{\mu}) d\psi^{1}_0 \!\ldots d\psi^{\mathcal{N}}_0,
\end{equation}
where $\mathcal{N}$ is the number of angular variables arising from the set $(q_1^{\mu}\ldots q_N^{\mu},p_{1\mu} \ldots p_{N\mu})$.  This expression gives the average of $F$ over one cyclic motion on all the zeroth-order tori.  Then, using the fact that $\langle \partial S_n/\partial \psi^{\mu}_0 \rangle = 0$, we find
\begin{align}
E_1(J_{\mu}) &= \langle H_1 \rangle \nonumber \\
E_2(J_{\mu}) &= \f{1}{\omega^{\mu}_0} \LB \left< \f{\partial H_1}{\partial J_{\mu}} \right> \langle H_1 \rangle - \left< \f{\partial H_1}{\partial J_{\mu}} H_1 \right> \RB \nonumber \\
&\quad + \f{1}{2} \f{1}{\omega^{\mu}_0} \f{1}{\omega^{\nu}_0} \f{\partial^2 H_0}{\partial J_{\mu} \partial J_{\nu}} \LB \langle H_1^2 \rangle - \langle H_1 \rangle^2 \RB.
\end{align}
Substituting these expressions in Eq.~(\ref{eq:E-S}), yields the differential equations that define the functions $S_n$. For example, for $S_1$
\begin{equation}
\omega^{\mu}_0 \f{\partial S_1}{\partial \psi^{\mu}_0} = \langle H_1 \rangle - H_1. \label{eq:S1-diffeq}
\end{equation}
The equations for the $S_n$ are readily solved by introducing the Fourier transform
\begin{equation}
S_n(\psi^{\mu}_0,J_{\mu}) = \sum_{m_1=-\infty}^{\infty} \cdots \sum_{m_\mathcal{N}=-\infty}^{\infty} S_n(m_{\mu},J_{\mu}) e^{i m_{\mu} \psi^{\mu}_0}.
\end{equation}
For $S_1$ one obtains
\begin{equation}
S_1(m_{\mu},J_{\mu}) = \f{K_1(m_{\mu},J_{\mu})}{i m_{\mu} \omega^{\mu}_0}, \label{eq:S1-soln}
\end{equation}
where $K_1(m_{\mu},J_{\mu})$ is the Fourier transform of the right-hand side of Eq.~(\ref{eq:S1-diffeq}).  This completes the formulation of CPT for our problem.  The above procedure can be carried to any order, although it becomes increasingly cumbersome at higher orders.  One must keep in mind the following important caveat.  CPT assumes the perturbation has a small effect on the zeroth-order dynamics.  But even if the perturbation is small in magnitude, if it is resonant its effect will not be.  The condition for resonance is that there exists some $m_{\mu}\pr=(m_1\pr,\ldots,m_\mathcal{N}\pr)$ such that the denominator in Eq.~(\ref{eq:S1-soln}) vanishes and $K_1(m_{\mu}\pr,J_{\mu})\neq 0$.  If a resonance occurs, it might still be possible to proceed by first making a canonical transformation that isolates the resonant variables \cite{kevorkian1996}.  If the dynamical equations for the resonant variables can be solved explicitly or numerically, CPT can be applied to the remaining degrees of freedom.  We shall see explicit examples of this in Sec.~\ref{sec:quench}.  In concluding this section, we remark that there is a related perturbation method, the Lie transformation method \cite{cary1981}, that is more convenient for performing explicit calculations to high order.

\section{\label{sec:eff:Ham} Effective Hamiltonians for reduced dynamics}

In Sec.~\ref{ssec:BBGKY:symplectic}, the symplectic structure of the BBGKY hierarchy was used to organize all of the degrees of freedom of $\rho$ into disjoint sets of canonically conjugate variables $\{q_n^{\mu},p_{n\mu}\}$, each associated with a particular level of the hierarchy.  The complete set of variables for level $n$ is $\{Q_n^{\mu},P_{n\mu}\} = \{q_1^{\mu},p_{1\mu}\} \cup \cdots \cup \{q_n^{\mu},p_{n\mu}\}$.  In this section, we address the following question:  for each level of the hierarchy, is there an effective Hamiltonian that generates the dynamics of the variables $\{Q_n^{\mu},P_{n\mu}\}$?  

Since the von Neumann equation has a Hamiltonian structure, there exists a Hamiltonian $H$ such that
\begin{align}
\dot{q}_n^{\mu} = \f{\partial H}{\partial p_{n\mu}}, \qquad
\dot{p}_{n\mu} = -\f{\partial H}{\partial q_n^{\mu}}. \label{eq:conjecture}
\end{align}
The ideal situation would be one in which there is a separation of variables, that is, the full Hamiltonian splits into terms
\begin{equation}
H = H_1(q_1^{\mu},p_{1\mu}) + H_2(q_2^{\mu},p_{2\mu}) + \cdots + H_N(q_N^{\mu},p_{N\mu}). \label{eq:separable}
\end{equation}
If this were the case, each $H_i$ would be a Hamiltonian function generating the dynamics of the reduced variables $\{q_i^{\mu},p_{i\mu}\}$, and $\sum_{i=1}^{n} H_i$ would be a Hamiltonian for the complete set $\{Q_n^{\mu},P_{\mu n}\}$.  Although an exact separation of variables will only occur in very special cases, the form in Eq.~(\ref{eq:separable}) might be a useful zeroth-order approximation for some systems.  The CPT of Sec.~\ref{sec:CPT} can be applied if the coupling between adjacent $\{q_n^{\mu},p_{n\mu}\}$ is weak.  But in general, we have to concede that the dynamics of the reduced variables $\{q_n^{\mu},p_{n\mu}\}$ might depend strongly on the variables of level $n+1$, since $\rho_{n+1}$ appears on the right-hand side of the equation of motion (\ref{eq:BBGKY-leveln}) for $\rho_n$.  

The general approach for obtaining effective dynamical equations for a set of reduced variables is to first derive an effective action by ``tracing out'' some (usually fast) degrees of freedom.  Then, the effective equations of motion are the Euler-Lagrange equations that follow from requiring the effective action to be stationary with respect to variations of the reduced variables.  This is not an exact approach, since an approximation is usually implied in directly tracing out some of the variables.  This is the point at which the possibility of using \textit{functionals} to close the hierarchy at a particular level comes in.  

\textbf{Conjecture} --- There exist Hamiltonian functionals $\mathcal{H}_n=\mathcal{H}_n[Q_n^{\mu},P_{n\mu}]$, depending also on the initial many-body state $\Psi_0$, such that the \textit{exact} dynamics of the complete set of reduced variables of any level $n$ are generated by the Hamilton equations
\begin{align}
\dot{Q}_n^{\mu} = \f{\partial \mathcal{H}_n}{\partial P_{n\mu}}, \qquad
\dot{P}_{n\mu} = -\f{\partial \mathcal{H}_n}{\partial Q_n^{\mu}}. \label{eq:conjecture}
\end{align}
The functionals $\mathcal{H}_n$ will generally depend on the entire history of the variables $Q_n^{\mu}=Q_n^{\mu}(t)$ and $P_{n\mu}=P_{n\mu}(t)$, which is referred to as \textit{memory dependence}.  Memory dependence arises when some subset of variables is eliminated in an exact way \cite{vankampen1985}.  For example, it appears explicitly as an integral over past times in the Nakajima-Zwanzig equation \cite{nakajima1958,zwanzig1960} for the reduced density matrix defined by tracing out the degrees of freedom of the environment.  Many other equations contain \textit{memory kernels} induced by the elimination of some set of variables.  In a similar way, the memory dependence in Eq.~(\ref{eq:conjecture}) is a consequence of eliminating the degrees of freedom $\{q_i^{\mu},p_{i\mu}\}$ with $i>n$.  We caution that $\mathcal{H}_n$ is a \textit{functional} and should not be interpreted as a Hamiltonian function over the reduced phase space $\{Q_n^{\mu}, P_{n\mu}\}$.

This approach to closing the BBGKY hierarchy is conceptually similar to  TD DFT, where the dynamical equations are closed in terms of the density. 
The exact time-dependent density can be calculated by propagating a set of single-particle Schr\"odinger equations with an effective potential $v_s=v_s(r,t)$, called the Kohn-Sham potential, which is a memory-dependent functional of the density.  TD DFT has a rigorous foundation, the Runge-Gross theorem \cite{runge1984}.  The existence of Hamiltonian functionals $\mathcal{H}_n$ that reproduce the exact dynamics of the reduced variables $\{Q_n^{\mu},P_{n\mu}\}$ remains a conjecture.  

In contrast to TD DFT, the present approach is not based on a one-to-one mapping between generalized potentials and generalized densities.  The results of Sec.~\ref{ssec:BBGKY:symplectic} regarding the symplectic geometry of the BBGKY hierarchy suggest that it might be possible to find a purely geometric proof of the existence of the $\mathcal{H}_n$.  However, even if the existence of exact $\mathcal{H}_n$ can be proven, we are still faced---as in TD DFT---with the problem of devising suitable functional approximations.  The Hamiltonian formulation of the BBGKY hierarchy is likely to serve as a springboard for introducing novel approximations.  Finally, we remark that this approach is more general than TD DFT or any of its extensions, which are limited to time-independent two-body interactions.  This excludes quantum quenches, such as the one studied in Sec.~\ref{sec:quench}, where the interaction is changed in time.

Let us pause and consider an example of what $\mathcal{H}_n$ might look like.  Consider a one-dimensional system of $N$ particles in a time-dependent state $| \Psi \rangle = | \Psi(t) \rangle$.  Suppose that the mean position of all particles, defined by $x = \f{1}{N} \sum_{i=1}^N \langle \Psi | x_i | \Psi \rangle$, undergoes approximately simple harmonic motion with slowly changing amplitude and frequency
\begin{equation}
x(t) = A(t) \cos \int^t \omega(t\pr) dt\pr.
\end{equation}
Take $x$ as representative of the $Q_1^{\mu}$.  An effective Hamiltonian that generates motion of this form is
\begin{equation}
\mathcal{H}_1 = \f{1}{2m} p^2 + k(t) x^2,
\end{equation}
where $p$ is the momentum conjugate to $x$.  The effective spring constant can be interpreted as a functional $k([x,p],t)$ that depends on the history of $x=x(t)$ and $p=p(t)$ for all $t\pr \leq t$.  This memory dependence accounts for the collective effect of all the degrees of freedom that have been eliminated.  If the changes in $\omega$ are slow, the action $J=\f{1}{2\pi} \oint p dx = E/\omega$ is an adiabatic invariant.

For the remainder of this section, let us focus on the closure of the BBGKY hierarchy in terms of the canonically conjugate variables $\{q_1^{\mu},p_{1\mu}\}$, which comprise all orbital degrees of freedom as well as the occupation numbers $n_k$ and their conjugate phases $\alpha_k$.  The $\alpha_k$ are important for generating the dynamics of the $n_k$ \cite{requist2011,requist2010,giesbertz2010b,giesbertz2012}.  The coupled $(\alpha_k,n_k)$-dynamics has been studied in linear response \cite{giesbertz2010b,giesbertz2010c,giesbertz2012} and in real time \cite{requist2011}.  In Ref.~\onlinecite{giesbertz2010b}, a TD DFT-like approach was introduced in which the dynamical equations are closed in terms of variables that are equivalent to the set $\{q_1^{\mu},p_{1\mu}\}$.  The equations of motion, derived from a stationary action principle, are a set of effective single-particle Schr\"odinger equations coupled to dynamical equations for the $n_k$.  
%  In concluding this section, we comment on the challenge of introducing functional approximations in terms of $\{q_1^{\mu},p_{1\mu}\}$.

One of the difficulties in devising functional approximations in terms of the variables $\{q_1^{\mu},p_{1\mu}\}$ is dealing with the $\alpha_k$ or $\zeta_k$ phase dependence.  The phases are often quite sensitive to the details of the dynamics and \textit{vice versa}. For example, the phases jump rapidly by $\pi$ whenever the $n_k$ approach the boundaries of the interval $[0,1]$; this changes the sign of $\dot{n}_k$ and maintains the Pauli principle  \cite{requist2011}.  One has little intuition what form the phase dependence should take.  A partial solution to this problem comes from realizing that the phases have a geometric significance as explained in Sec.~\ref{ssec:BBGKY:geometric-phase}.  Geometric phases appear as \textit{action integrals} associated with cyclic evolutions of the set $\{q_1^{\mu},p_{1\mu}\}$.  Action integrals are important because they often have a transparent physical meaning.

Expressing functional approximations for $\mathcal{H}_1[q_1^{\mu},p_{1\mu}]$ in terms of approximate action-angle variables appears to be a promising approach.  Although exact action-angle variables do not generally exist, in many cases it will be possible to transform to optimal action-angle variables $\{\psi^{\mu},J_{\mu}\}$ for which the $J_{\mu}$ are slowly varying.  The exact functional will generally contain angle dependence that cannot be eliminated by transforming to optimal action-angle variables, yet it will be significantly weaker if the $J_{\mu}$ are slowly varying since $\dot{J}_{\mu} = -\partial \mathcal{H}_1/\partial \psi^{\mu}$.  Moreover, memory dependence has been found to take a simple form when the system is integrable or possesses adiabatic invariants, i.e.~approximate constants of the motion \cite{requist2011}.  

Given that the equations of motion are expressed in the form of the classical Hamilton equations, it might be profitable to study further the relationship between angle dependence, memory dependence and geometric phase.  All geometric phases carry a form of memory dependence because they are nonintegrable phases.    For integrable classical systems, the Hannay angle \cite{hannay1985} is a geometric phase that carries a memory of where the system went.  If the system is not integrable but admits a fast/slow separation of variables, an effective functional governing the slow variables can be found.  Such functionals lose short-term memory dependence through rapid oscillations of the fast variables \cite{requist2011}.  Within linear response, memory-dependent functional approximations have been derived from reference systems such as the electron gas (see for example Refs.~\onlinecite{gross1985} and \onlinecite{vignale1996}).  This is much harder to do in strongly nonlinear regimes.  One potential strategy for quantifying the memory dependence of a reference system in a nonlinear regime would be to parametrize the hysteresis loops corresponding to cyclic motions of the reduced variables.

\section{\label{sec:quench} Interaction quench in a finite Hubbard chain}

In this section, we apply CPT within the Hamiltonian formulation of the BBGKY hierarchy to describe quench dynamics in a Hubbard chain initially undergoing density oscillations.  For $t<0$, the Hubbard parameter $U=U_i$ is chosen to be much less than the hopping $V$.  At time $t=0$, $U$ is suddenly increased (quenched) to a large value $U_f \gg V$.  The dynamics in the weakly and strongly interacting regimes are qualitatively different.  We shall find that in both regimes we can make a separation of fast/slow degrees of freedom.  Interestingly, the identity of the fast and slow variables is interchanged by the quench.

For the sake of clarity, we consider the simplest possible Hubbard chain: one with just two sites and two electrons.  This model has been studied previously in a different dynamical scenario, namely under a linear ramping of the bias between the two sites \cite{requist2010,requist2011}.  Despite its simplicity, the model displays nontrivial dynamics.  Some of the qualitative conclusions that we can draw are applicable to all finite Hubbard chains.  Due to the reduced dimension of the Hilbert space, we will be able to carry out CPT fully analytically.  The dynamics is in fact integrable in each of the two regimes, but we shall not derive the exact solution as our aim is only to illustrate the application of CPT to nonequilibrium quantum dynamics.  For two electrons the BBGKY hierarchy of course truncates at second order.  Therefore, our analysis is not an ideal example of the approach outlined in Sec.~\ref{sec:CPT}.  Nevertheless, it is representative of the general structure of the problem.  The Hamiltonian is 
\begin{equation}
H = -\f{V}{2} \sum_{\sigma} (c_{1\sigma}^{\dag} c_{2\sigma} + c_{2\sigma}^{\dag} c_{1\sigma}) + U(t) (n_{1\uparrow} n_{1\downarrow} + n_{2\uparrow} n_{2\downarrow}). \label{eq:Hamiltonian:Hubbard}
\end{equation}
In this model, the density is represented by the variable $Z \equiv \sum_{\sigma} \langle n_{1\sigma}-n_{2\sigma} \rangle/2$.  The initial condition at $t_0=-46$ is taken to be a state with $Z \neq 0$, so that in the regime before the quench the density undergoes persistent oscillations.  Since there are no spin-flip terms in Eq.~(\ref{eq:Hamiltonian:Hubbard}), $S^2$ and $S_z$ are conserved and we consider only the sector of spin-singlet states with $S_z=0$.

Before beginning our analysis of the two dynamical regimes, let us identify a complete set of canonically conjugate variables.  First, note that $\rho_1$ can be mapped to a vector $\vec{\rho}$ lying within the so-called Bloch ball, defined by $|\vec{\rho}| \leq 1$, through the equation
\begin{equation}
\rho_1 = I + \vec{\rho}\cdot\vec{\sigma}.
\end{equation}
The north pole of the Bloch sphere corresponds to having both electrons in site $1$.  Due to correlations, the modulus $|\vec{\rho}|$ can be less than 1.  Let $\varphi$ be the azimuthal angle of the vector $\vec{\rho}$.  $(q^1,p_1) \equiv (\varphi, Z)$ are a pair of conjugate variables, cf.~Eq.~(\ref{eq:Hamilton-canonical}).  The other pair is $(q^2,p_2) \equiv (2\zeta, A/2)$, where $\zeta$ is a phase degree of freedom of $\rho_2$ and $A\equiv |\vec{\rho}| = (n_a-n_b)/2$; $n_a$ and $n_b$ ($n_a\geq n_b$) are the eigenvalues of $\rho_1$.  The BBGKY equations are equivalent to the Hamilton equations
\begin{align}
\dot{\varphi} &= \frac{\partial H}{\partial Z}  & \dot{\zeta} &= \frac{\partial H}{\partial A} \nonumber \\
\dot{Z} &= -\frac{\partial H}{\partial \varphi}& \dot{A} &= -\frac{\partial H}{\partial \zeta}, \label{eq:2HM:Hamilton}
\end{align}
with the Hamiltonian function
\begin{align}
H &= -V \sqrt{A^2-Z^2} \cos\varphi + \f{U}{2} \f{A^2+Z^2}{A^2} \nonumber\\
&\quad -\f{U}{2}  \f{A^2-Z^2}{A^2} B \cos2\zeta,
\end{align}
where, for the sake of brevity in the following results, we have defined $B=\sqrt{1-A^2}$.

\begin{figure}[t!]
\includegraphics[width=1.0\columnwidth]{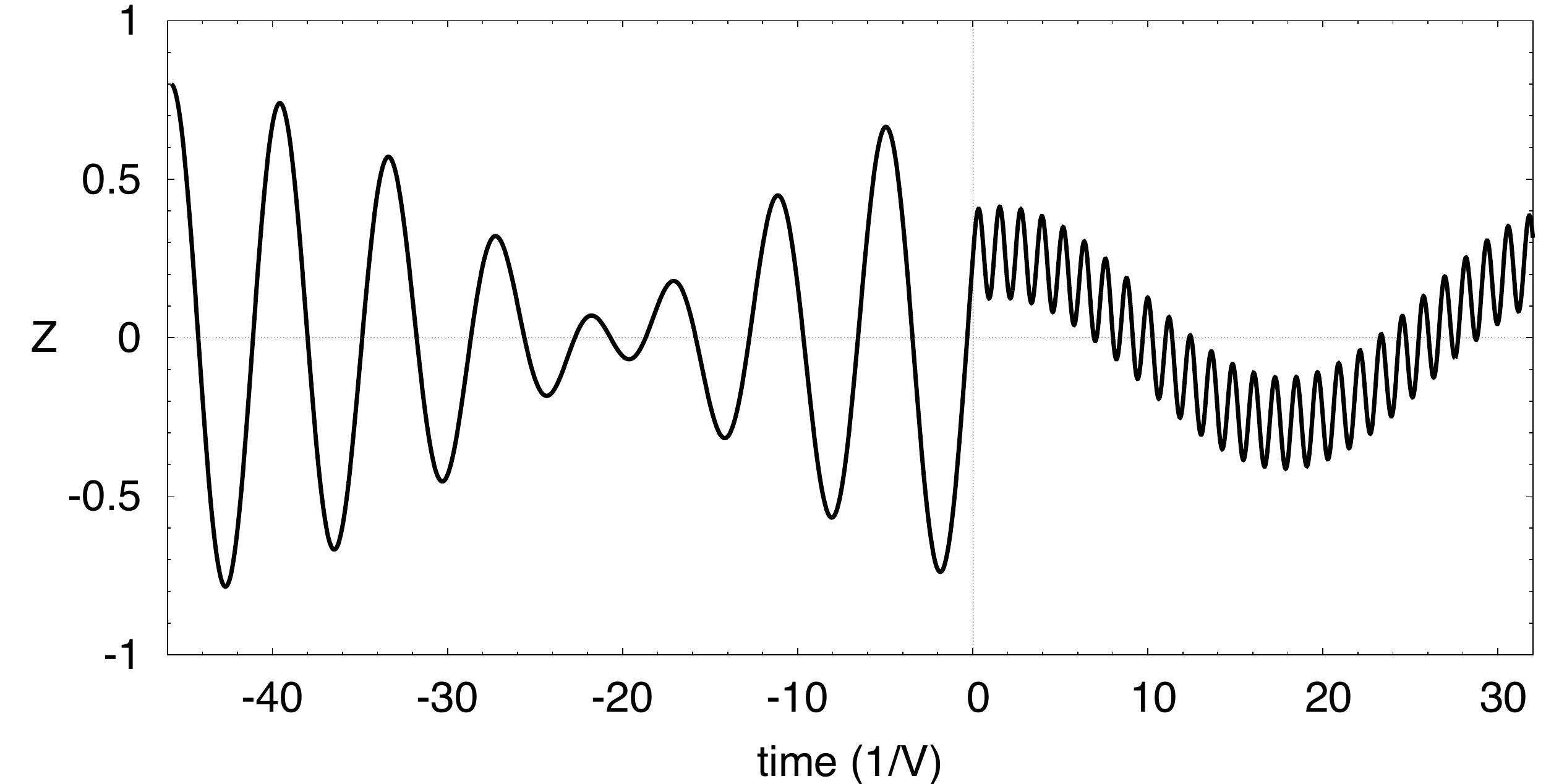}
\caption{\label{fig:Z-fulltime} Dynamics of the density $Z$; quench at $t=0$.}
\end{figure}

\begin{figure}[t!]
\includegraphics[width=1.0\columnwidth]{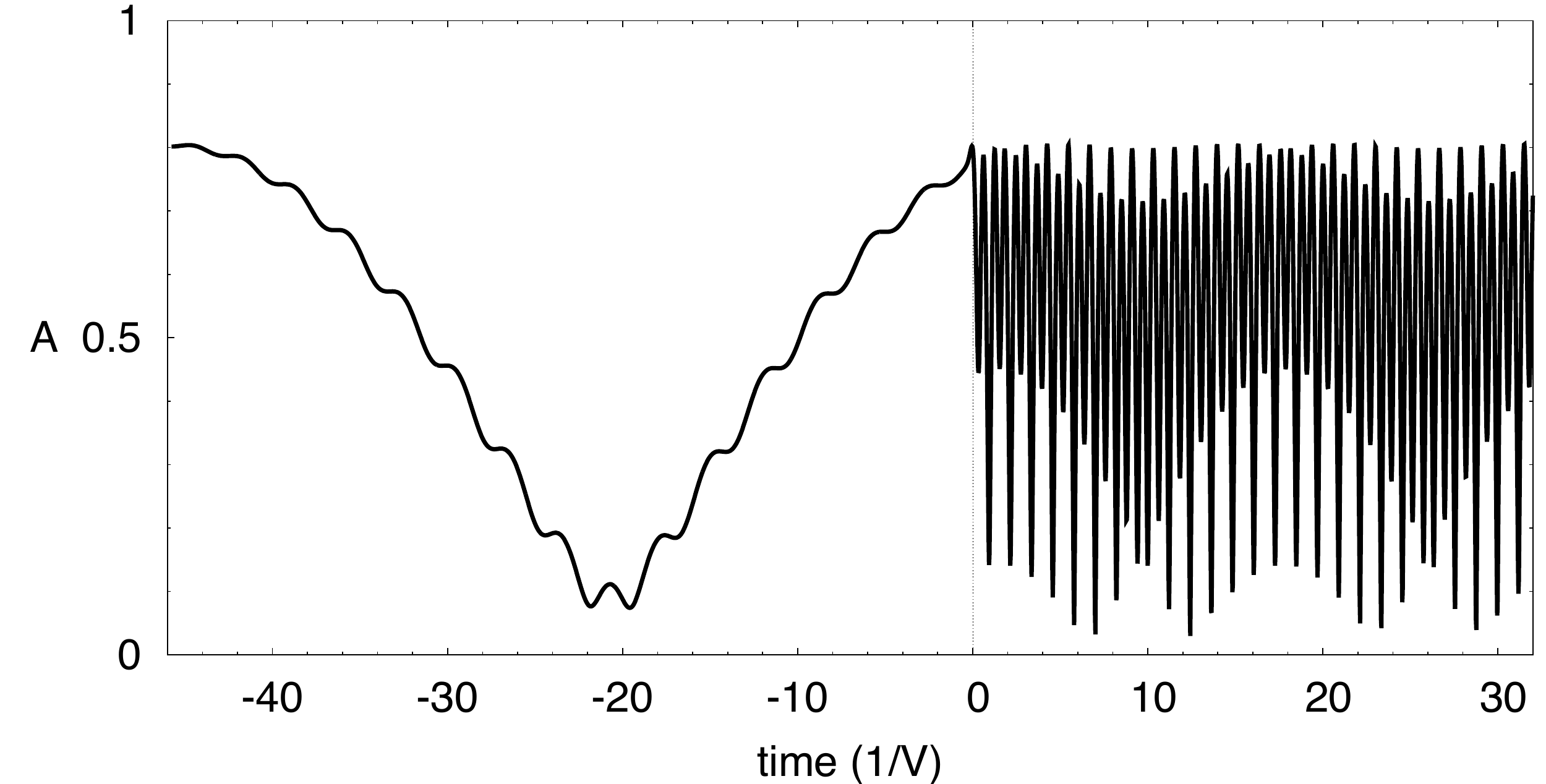}
\caption{\label{fig:A-fulltime} Dynamics of the correlation $A$; quench at $t=0$.}
\end{figure}

Figures~\ref{fig:Z-fulltime} and \ref{fig:A-fulltime} show the full time evolution of the variables $Z$ and $A$.  Before the quench the density undergoes persistent harmonic oscillations of period $2\pi/V$ with beating on the longer time scale $2\pi/U_i$.  The parameters are $V=2$, $U_i=1/4$ and $U_f=10$ in arbitrary units.  To understand why the oscillations in $Z$ are collapsing, we look at the behavior of the variable $A$, which is directly related to the occupation numbers.  Physically, $A$ is a measure of the correlation of the system; $A=1$ corresponds to an uncorrelated state while $A=0$ corresponds to the maximally correlated state.  
In Fig.~\ref{fig:A-fulltime}, we see that the point of collapse of the oscillations coincides with the minimum of $A$.  This means that the kinetic energy of the oscillations has been converted into internal correlation energy.  The build up and decay of this correlation occurs periodically with a frequency $U_i$ set by the interaction strength.  

We have a clear separation of time scales: before the quench, the rapidly oscillating density is the fast variable and the slowly oscillating internal correlation is the slow variable.  After the quench, when the Hubbard interaction has been increased to $U_f \gg V$, the situation is inverted.  The internal correlation undergoes rapid oscillations and becomes the fast variable.  The density displays low frequency oscillations, together with higher frequency overtones, and therefore represents the slow variable.

\subsection{\label{ssec:quench:before} Before the quench: weak interaction regime}

In the weakly interacting regime before the quench, we treat the interaction terms as the perturbation $H_1$.  The hopping (kinetic) terms give the zeroth-order Hamiltonian $H_0$.  Since the dynamics are integrable for $H_0$, there exist zeroth-order AA variables.  The first action variable can be calculated as 
\begin{align}
I_1 &= \f{1}{2\pi} \oint Z d\varphi = Q_1 - Q_2, \label{eq:before:I1}
\end{align}
where we have introduced the constants of the motion $Q_1 = A$ and $Q_2 = H_0/V = \sqrt{A^2-Z^2} \cos\varphi$.  The integral in Eq.~(\ref{eq:before:I1}) has been evaluated using the residue theorem. The second action variable is  
\begin{align}
I_2 &= \f{1}{2\pi} \oint A d\zeta = \f{Q_1}{2}.
\end{align}
The angle variables can be obtained from the Hamilton characteristic function $W = W_1(q_1,Q) + W_2(q_2,Q)$ with  
\begin{align}
W_1 &= \int \gamma d\varphi \nonumber \\
&= Q_1 \sin^{-1} \f{Q_1}{\sqrt{Q_1^2-Q_2^2}} \sin\varphi \nonumber \\
&- Q_2 \sin^{-1} \f{Q_2}{\sqrt{Q_1^2-Q_2^2}} \tan\varphi.
\end{align}
and $W_2 = A\zeta$.  The angle variable corresponding to $I_1$ is 
\begin{align} 
\phi_1 = \f{\partial \tilde{W}}{\partial I_1} = \sin^{-1} \f{Q_2}{\sqrt{Q_1^2-Q_2^2}} \tan\varphi.
\end{align}
Here, the function $\tilde{W}$ is obtained from $W$ by substituting $Q_{\mu}=Q_{\mu}(I)$.  The angle variable corresponding to $I_2$ is
\begin{align}
\phi_2 = \f{\partial \tilde{W}}{\partial I_2} = 2\zeta_1 - 2\phi_1,
\end{align}
where we have defined $\zeta_1 = \zeta - \zeta_0$ with
\begin{align}
\zeta_0 &= -V \int d\phi_1 \LB \f{d\phi_1}{dt} \RB^{-1} \f{Q_1 Q_2}{Q_1^2 \sin^2\phi_1 + Q_2^2 \cos^2\phi_1} \nonumber \\
&= -\sin^{-1} \f{Q_1}{\sqrt{Q_1^2-Q_2^2}} \sin\varphi
\end{align}
giving the zeroth-order dynamics of $\zeta$.  Expressed in terms of the AA variables, the zeroth-order Hamiltonian function is
\begin{equation}
H_0 = V I_1 - 2V I_2.
\end{equation}
To zeroth order, $A=const$ and the dynamics consists of simple harmonic oscillations of the density expressible as
\begin{equation}
Z = \sqrt{Q_1^2-Q_2^2} \cos\phi_1
\end{equation}
with $\phi_1 = V t + const$.

Now we would like to carry out CPT with respect to the perturbation $H_1 = H - H_0$ following the procedure outlined in Sec.~\ref{sec:CPT}.  It turns out that this is a case where CPT cannot be applied naively because the perturbation is resonant.  The factor $m_1 \omega_1 + m_2 \omega_2$ vanishes for the integers $m_1 = 2$ and $m_2 = 1$, while the corresponding Fourier component of $K_1 = \langle H_1 \rangle - H_1$ is nonzero; cf. Eq.~(\ref{eq:S1-soln}).  Fortunately, there is a way around this problem (see for example Ref.~\onlinecite{kevorkian1996}).  The solution is to make a canonical transformation that isolates the resonant variable $\phi\pr_1 = (2\phi_1 + \phi_2)/2$.  Since the dynamical equations for $\phi\pr_1$ and its conjugate variable $I\pr_1$ --- the \textit{resonant pair} --- decouple from the remaining variables, they can be solved by quadrature.  Then, CPT can proceed as usual.  Physically, the consequence of the resonance is that the corrections to the zeroth-order action variables $I_1$ and $I_2$ are of order $\mathcal{O}(1)$ rather than $\mathcal{O}(U/V)$ as we would have expected if there had been no resonance.  This is why the quantity $A$, shown in Fig.~\ref{fig:A-fulltime}, changes by $\mathcal{O}(1)$ even though the zeroth-order result predicts $A=const$.

Following Ref.~\onlinecite{kevorkian1996}, we set
\begin{align}
\phi\pr_1 &= \phi_1 + \f{1}{2} \phi_2 &  I\pr_1 &= I_1 \nonumber \\
\phi\pr_2 &= \phi_2 &  I\pr_2 &= -\f{1}{2} I_1 + I_2. \label{eq:isolating}
\end{align}
The generating function of this transformation is
\begin{equation}
S(\phi_{\mu}, I_{\mu}\pr) = \Big( \phi_1 + \f{1}{2} \phi_2\Big) I\pr_1 + \phi_2 I\pr_2.
\end{equation}
The new Hamiltonian is split up as follows:
\begin{equation}
H(\phi\pr_{\mu},I\pr_{\mu}) = H_0 + \langle H_1\rangle + H_{1c} + H_{1s} + H_2, \label{eq:H:split}
\end{equation}
where 
\begin{align*}
H_0 &= -2 V I\pr_2 \\
\langle H_1 \rangle &= \f{U}{2} + \f{U}{4} \f{I\pr_1(I\pr_1 + 4 I\pr_2)}{(I\pr_1 + 2 I\pr_2)^2} \\
H_{1c} &= \f{UB}{4} \f{I\pr_1(I\pr_1+4 I\pr_2)}{(I\pr_1 + 2 I\pr_2)^2} \cos 2\phi\pr_1 \\
H_{1s} &= -\f{U}{4} \f{Q_2^2 - Q_1^2}{Q_1^2} \cos2\phi_1 - \f{UB}{2} \f{Q_2}{Q_1} \sin2\phi_1 \sin2\phi\pr_1 \\
&\quad - \f{UB}{4} \f{Q_2^2 + Q_1^2}{Q_1^2} \cos2\phi_1 \cos2\phi\pr_1.
\end{align*}
The critical term $H_{1c}$ is defined as the part of $H_1 - \langle H_1 \rangle$ that does not vanish upon averaging over $\phi\pr_2$; it depends only on the resonant angle $\phi\pr_1$.  $H_{1s} = H_1 - \langle H_1 \rangle - H_{1c}$ is the noncritical remainder.  For brevity, we have expressed $H_{1s}$ in terms of $Q_{\mu}$ instead of $I\pr_{\mu}$.

To carry out CPT to second order in $H_1$, the next step is to define the near-identity canonical transformation $(\phi\pr_{\mu},I\pr_{\mu}) \rightarrow (\psi_{\mu},J_{\mu})$ that takes into account 
\begin{figure}[t!]
\includegraphics[width=1.0\columnwidth]{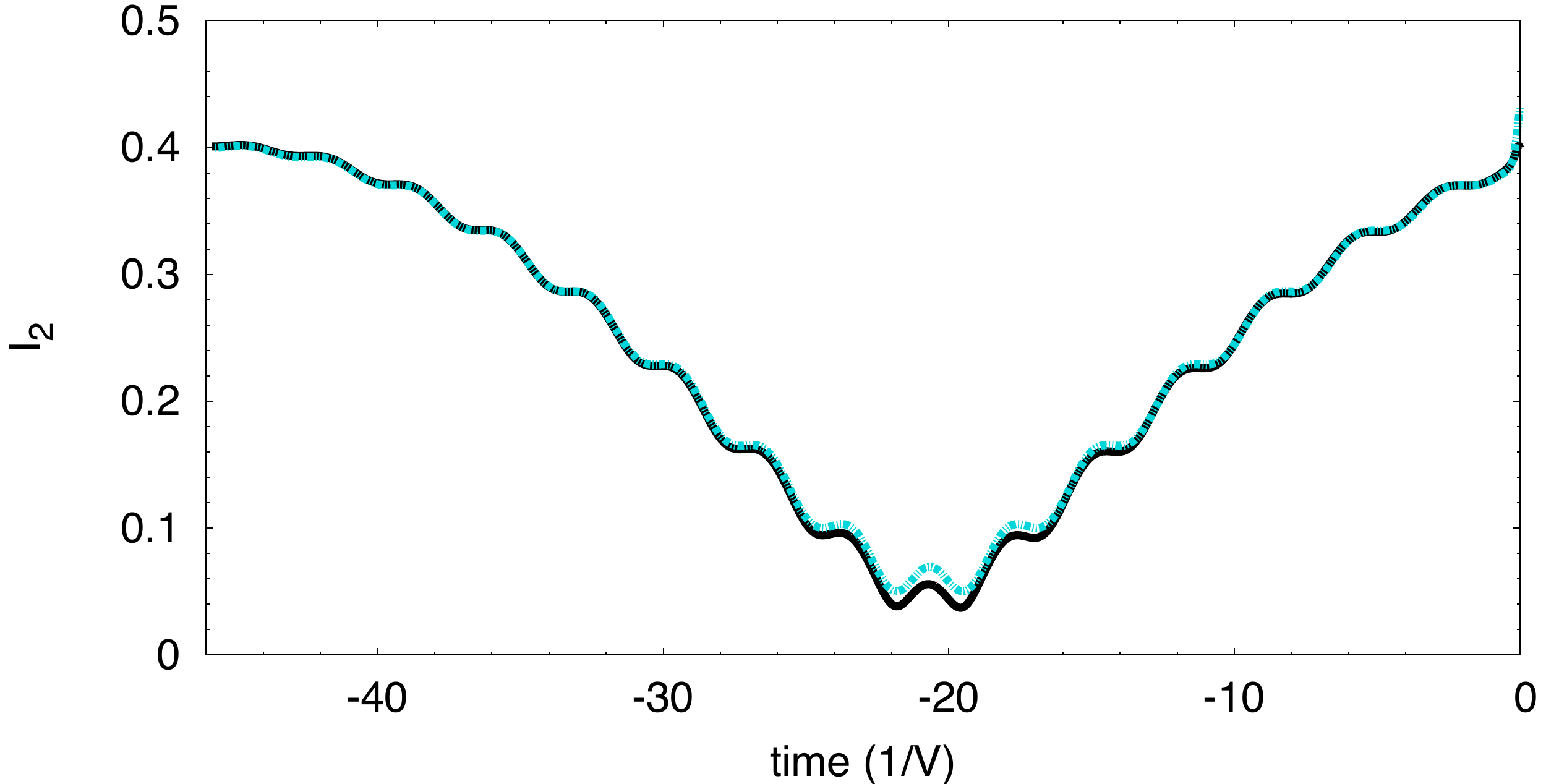}
\caption{\label{fig:I2-before} (Color online) Exact (black solid) and first-order (cyan dashed) dynamics of $I_2=A/2$ before the quench.  Approximate result from Eq.~(\ref{eq:AA-before-approx}).}
\end{figure}
\begin{figure}[t!]
\includegraphics[width=1.0\columnwidth]{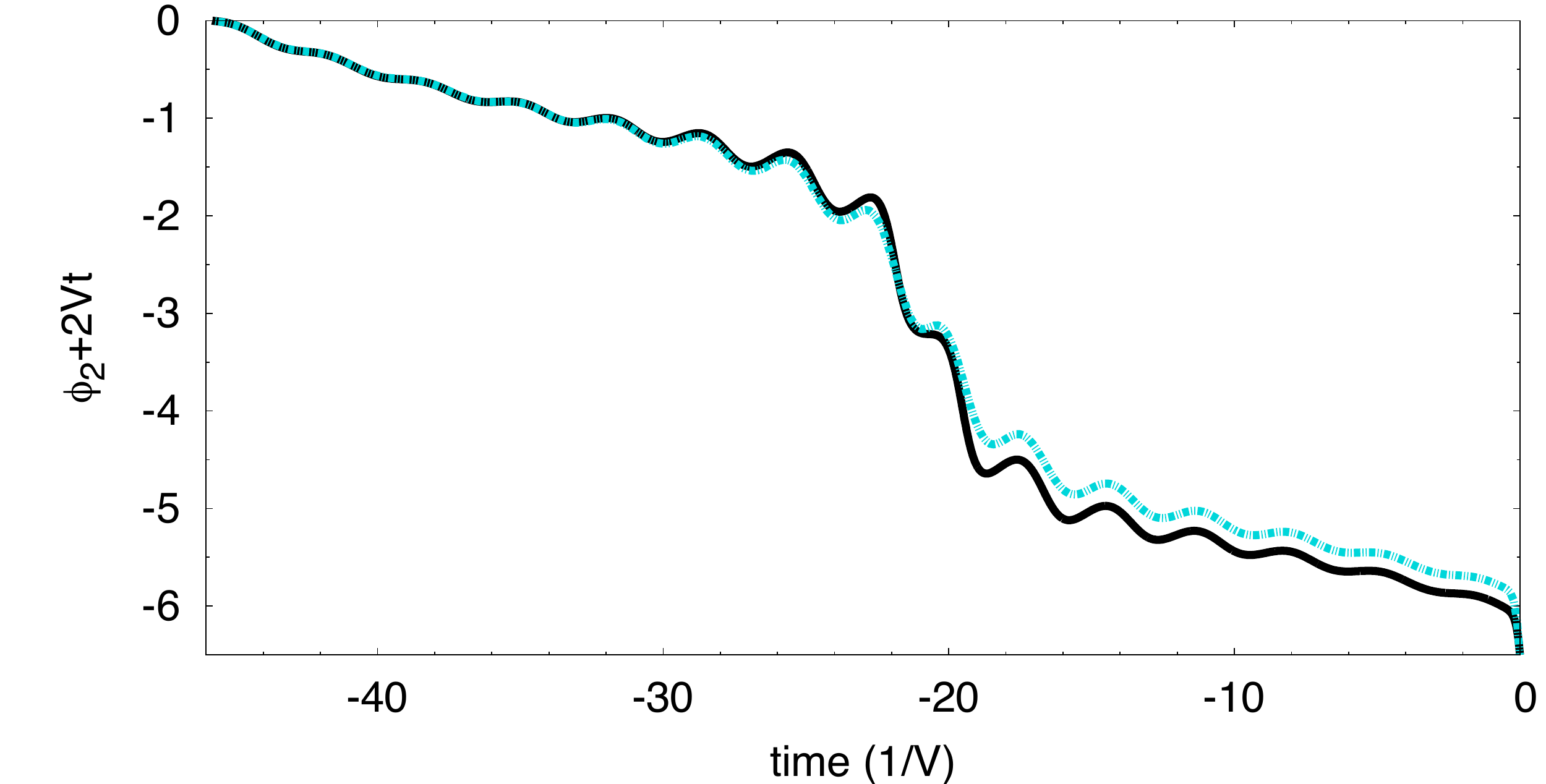}
\caption{\label{fig:phi2-before} (Color online) Exact (black solid) and first-order (cyan dashed) dynamics of $\phi_2+2Vt$ before the quench.  Approximate result from Eq.~(\ref{eq:AA-before-approx}).}
\end{figure}
only the noncritical part $H_{1s}$ of the perturbation.  The generating function is $F(\phi\pr_{\mu},J_{\mu}) = \phi\pr_{\mu} J_{\mu} + F_1(\phi\pr_{\mu},J_{\mu})$, where 
\begin{equation}
\omega\pr_1 \f{\partial F_1}{\partial \phi\pr_1} + \omega\pr_2 \f{\partial F_1}{\partial \phi\pr_2} = -H_{1s} \label{eq:F1-diffeq}
\end{equation}
with $\omega\pr_1 = \omega_1 + \f{1}{2} \omega_2 = 0$ and $\omega\pr_2 = \omega_2 = -2 V$.  Now, since we have removed the critical term $H_{1c}$ from the right-hand side of Eq.~(\ref{eq:F1-diffeq}), no divergences appear.  Integrating Eq.~(\ref{eq:F1-diffeq}), we find
\begin{align}
F_1 &= -
\f{U}{8V} \f{Q_1^2 - Q_2^2}{Q_1^2} \sin2\phi_1 - \f{UB}{4V} \f{Q_2}{Q_1} \cos2\phi_1 \sin2\phi\pr_1 \nonumber \\
&\quad + \f{UB}{8V} \f{Q_1^2  + Q_2^2}{Q_1^2} \sin2\phi_1 \cos2\phi\pr_1 . 
\end{align} 
We have suppressed an arbitrary function $f_1(J_{\mu})$.  Note that $F_1$ is a function of $J_{\mu}$ obtained by evaluating $Q_{\mu}(I\pr_{\nu})$ for $I\pr_{\nu}=J_{\nu}$.  Now the partially-averaged Hamiltonian in the $(\psi_{\mu},J_{\mu})$ variables depends only on $\psi_1$.  It is 
\begin{equation}
\mathcal{H}(\psi_{\mu},J_{\mu}) = \mathcal{H}_0(J_{\mu}) + \mathcal{H}_1(\psi_1,J_{\mu}) + \mathcal{H}_2(J_{\mu}), \label{eq:H}
\end{equation}
where
\begin{align}
\mathcal{H}_0 &= \langle H_0 \rangle \nonumber \\ 
\mathcal{H}_1 &= \langle H_1 \rangle + H_{1c}  \nonumber \\
\mathcal{H}_2 &= \langle H_2 \rangle + \f{\partial f_1}{\partial t} - \langle G \rangle \label{eq:H-terms}
\end{align}
and the $\mathcal{H}_i$ are obtained by evaluating the corresponding terms of $H(\phi\pr_{\mu},I\pr_{\mu})$ for $\phi\pr_{\mu} = \psi_{\mu}$ and $I\pr_{\mu}=J_{\mu}$.  The function $G$ is
\begin{equation}
G = \f{1}{2} \,\f{\partial^2 \langle H_0 \rangle}{\partial I\pr_{\mu} \partial I\pr_{\nu}} \: \f{\partial F_1}{\partial \phi\pr_{\mu}} \: \f{\partial F_1}{\partial \phi\pr_{\nu}} + \f{\partial \langle H_0 \rangle}{\partial I\pr_{\nu}} \: \f{\partial^2 F_1}{\partial \phi\pr_{\nu} \partial I\pr_{\mu}} \:\f{\partial F_1}{\partial \phi\pr_{\mu}}, 
\end{equation}
where, \textit{after} the differentiation is performed, the right-hand side is evaluated for $\phi\pr_{\mu} = \psi_{\mu}$ and $I\pr_{\mu}=J_{\mu}$.  In Eq.~(\ref{eq:H}), we have kept the average terms to order $U^2$ but not the oscillatory terms.  The arbitrary function $f_1(J_{\mu})$ that appeared above can be chosen so that $\mathcal{H}_2$ vanishes.  Then, we have
\begin{align}
\dot{J}_1 &= -\f{\partial \mathcal{H}_{1c}}{\partial \psi_1} \nonumber \\
&= \f{UB}{2} \f{J_1(J_1 + 4 J_2)}{(J_1 + 2 J_2)^2} \sin2\psi_1 \nonumber \\
\dot{\psi}_1 &= \omega\pr_1(J_{\mu}) + \f{\partial \mathcal{H}_1}{\partial J_1}  \nonumber \\
&= 2U \f{J_2^2}{(J_1+2 J_2)^3} \LB 1 + \f{\cos2\psi_1}{B} \RB \nonumber\\
&\quad - \f{U}{4} \f{\cos2\psi_1}{B} \LB \f{4 J_2^2}{J_1 + 2J_2} + J_1 + 2J_2 \RB  \label{eq:system1} \\[0.2cm]
\dot{J}_2 &= \mathcal{O}(U^2) \nonumber \\
\dot{\psi}_2 &= \omega\pr_2(J_{\mu}) + \f{\partial \mathcal{H}_1}{\partial J_2} \label{eq:system2}.
\end{align}
In these dynamical equations, oscillatory terms of $\mathcal{O}(U^2)$ that vanish upon averaging over $\psi_1$ and $\psi_2$ have been neglected.  The pair $(\psi_1,J_1)$ obey the first-order system Eq.~(\ref{eq:system1}) in which $J_2$ is regarded as constant. 
If we find the solution of these equations, then it is straightforward to calculate $\psi_2$ by quadrature.  For this we need 
\begin{align}
\f{\partial \mathcal{H}_1}{\partial J_2} &= -2U \f{J_1 J_2}{(J_1 + 2 J_2)^3} \LB 1 + \f{\cos2\psi_1}{B} \RB \nonumber \\
&\quad - \f{U}{2} \f{J_1^2}{J_1 + 2 J_2} \f{\cos2\psi_1}{B}. 
\end{align}
Then, by integration,
\begin{equation}
\psi_2 = \omega\pr_2 (t-t_0) + \int_{t_0}^t \f{\partial \mathcal{H}_1}{\partial J_2} ds .
\end{equation}
Finally, we use the following inverse transformations to return to the original variables:
\begin{align}
I_1 &= J_1 + \f{\partial F_1}{\partial \phi\pr_1} \nonumber \\
I_2 &= J_2 + \f{1}{2} J_1 + \f{\partial F_1}{\partial \phi\pr_2} + \f{1}{2} \f{\partial F_1}{\partial \phi\pr_1} \nonumber \\
\phi_1 &= \psi_1 - \f{1}{2} \psi_2 - \f{\partial F_1}{\partial J_1} + \f{1}{2} \f{\partial F_1}{\partial J_2} \nonumber \\
\phi_2 &= \psi_2 - \f{\partial F_1}{\partial J_2}. \label{eq:AA-before-approx}
\end{align}
In Figs.~\ref{fig:I2-before} and \ref{fig:phi2-before}, the approximate results for $I_2=A/2$ and $\phi_2+2Vt$ are compared with the exact results.  The approximate results for all other quantities, namely $I_1$ and $\phi_1$, are equally good.

\subsection{\label{ssec:quench:after} After the quench: strong interaction regime}

In the strongly interacting regime, the situation is inverted.  Here we treat the hopping terms as the perturbation $H_1$ and let the interaction terms be $H_0$.  Like the regime before the quench, the zeroth-order dynamics are integrable.  The action variables are (for uniformity we use the same symbols as in the previous section)
\begin{align}
I_1 &= \f{1}{2\pi} \oint Z d\varphi = Q_1 \nonumber \\
I_2 &= \f{1}{2\pi} \oint A d\zeta = \f{1}{2} (1-Q_1-Q_2)
\end{align}
where we have defined the constants $Q_1=Z$ and 
\begin{align}
Q_2 &= \f{H_0}{U} = \f{1}{2} \f{A^2+Z^2}{A^2} - \f{1}{2}  \f{A^2-Z^2}{A^2} B \cos2\zeta.
\end{align}
As previously, the angle variables are calculated from the Hamilton characteristic function.  Since the expressions are lengthy, we shall report only the result:  
\begin{align}
\phi_1 &= \f{\partial \tilde{W}}{\partial I_1} = \varphi_1 + \f{\phi_2}{2} \nonumber \\
\phi_2 &= \f{\partial \tilde{W}}{\partial I_2} = -2 \: \mathrm{sgn}(\dot{A}) \: \tan^{-1} \LB \f{A^2-A_1^2}{A_2^2-A^2} \RB^{1/2},
\end{align}
where $A_1$ and $A_2$ are defined below and we have set $\varphi = \varphi_0 + \varphi_1$ with
\begin{align}
\varphi_0 &= 2U \int d\phi_2 \LB \f{d\phi_2}{dt} \RB^{-1} \f{Z(1-Q_2)}{A^2-Z^2} \nonumber  \\
&= \tan^{-1} \left[ \LB \f{A_2^2-Z^2}{A_1^2-Z^2} \RB^{1/2} \tan \f{\phi_2}{2} \right].
\end{align}
The zeroth-order Hamiltonian function is
\begin{equation}
H_0 = U - U I_1 - 2U I_2. 
\end{equation}
To zeroth-order $Z=const$, and the dynamics is given by harmonic oscillations of $A^2$:
\begin{align}
A^2 &= A_1^2 \cos^2 \f{\phi_2}{2} + A_2^2 \sin^2 \f{\phi_2}{2} \nonumber \\
&= \f{1}{2} (A_1^2 + A_2^2) + \f{1}{2} (A_1^2 - A_2^2) \cos \phi_2
\end{align}
where $\phi_2 = -2U t + const.$  The turning points of the oscillations are
\begin{align}
A_{1,2}^2 = Q_1^2 + 2 (1-Q_2) \Big( Q_2 \mp \sqrt{Q_2^2-Q_1^2} \Big).
\end{align}

In the strongly interacting regime, the perturbation is nonresonant so we can apply CPT straightforwardly.  Let $S(\psi_{\mu0},J_{\mu1}) = \psi_{\mu0} J_{\mu1} + S_1(\psi_{\mu0},J_{\mu1})$ be the generating function of the transformation $(\psi_0,J_0) \rightarrow (\psi_1,J_1)$, where for the zeroth-order variables we have changed notation according to $(\psi_0,J_0) = (\phi,I)$.  The first-order part, $S_1$, is given by the differential equation
\begin{equation}
\omega_{10} \f{\partial S_1}{\partial \psi_{10}} + \omega_{20} \f{\partial S_1}{\partial \psi_{20}} = \langle H_1 \rangle - H_1, \label{eq:S1-diffeq2}
\end{equation}
which can be solved by Fourier transform as in Eq.~(\ref{eq:S1-soln}).  The only nonzero Fourier components are
\begin{align*}
K_{1,0} = K_{-1,0} &= \f{V}{2} \sqrt{\f{1-Q_2}{2}} \left[ \sqrt{ Q_2 + (Q_2^2-Q_1^2)^{1/2} } \right. \\ 
&\quad\left. -\mathrm{sgn}(Q_1) \sqrt{ Q_2 - (Q_2^2-Q_1^2)^{1/2} } \right]  \\
K_{1,-1} = K_{-1,1} &= -\f{V}{2} \sqrt{\f{1-Q_2}{2}} \left[ \sqrt{ Q_2 + (Q_2^2-Q_1^2)^{1/2} } \right. \\
&\quad\left. +\mathrm{sgn}(Q_1) \sqrt{ Q_2 - (Q_2^2-Q_1^2)^{1/2} } \right], 
\end{align*}
where the right-hand side is evaluated at $I_{\mu} = J_{\mu1}$.  Thus, for $S_1$ we obtain
\begin{equation*}
S_1(\psi_0,J_1) = \f{2K_{1,0}}{\omega_{10}} \sin\psi_{10} + \f{2K_{1,-1}}{\omega_{10}-\omega_{20}} \sin(\psi_{10}-\psi_{20}).
\end{equation*}
The first-order AA variables are
\begin{align}
\psi_{k1} &= \psi_{k0} + \f{\partial S_1}{\partial J_{k1}} \nonumber \\
J_{k0} &= J_{k1} + \f{\partial S_1}{\partial \psi_{k0}} \label{eq:AA-after-approximate}.
\end{align}

\begin{figure}[t!]
\includegraphics[width=1.0\columnwidth]{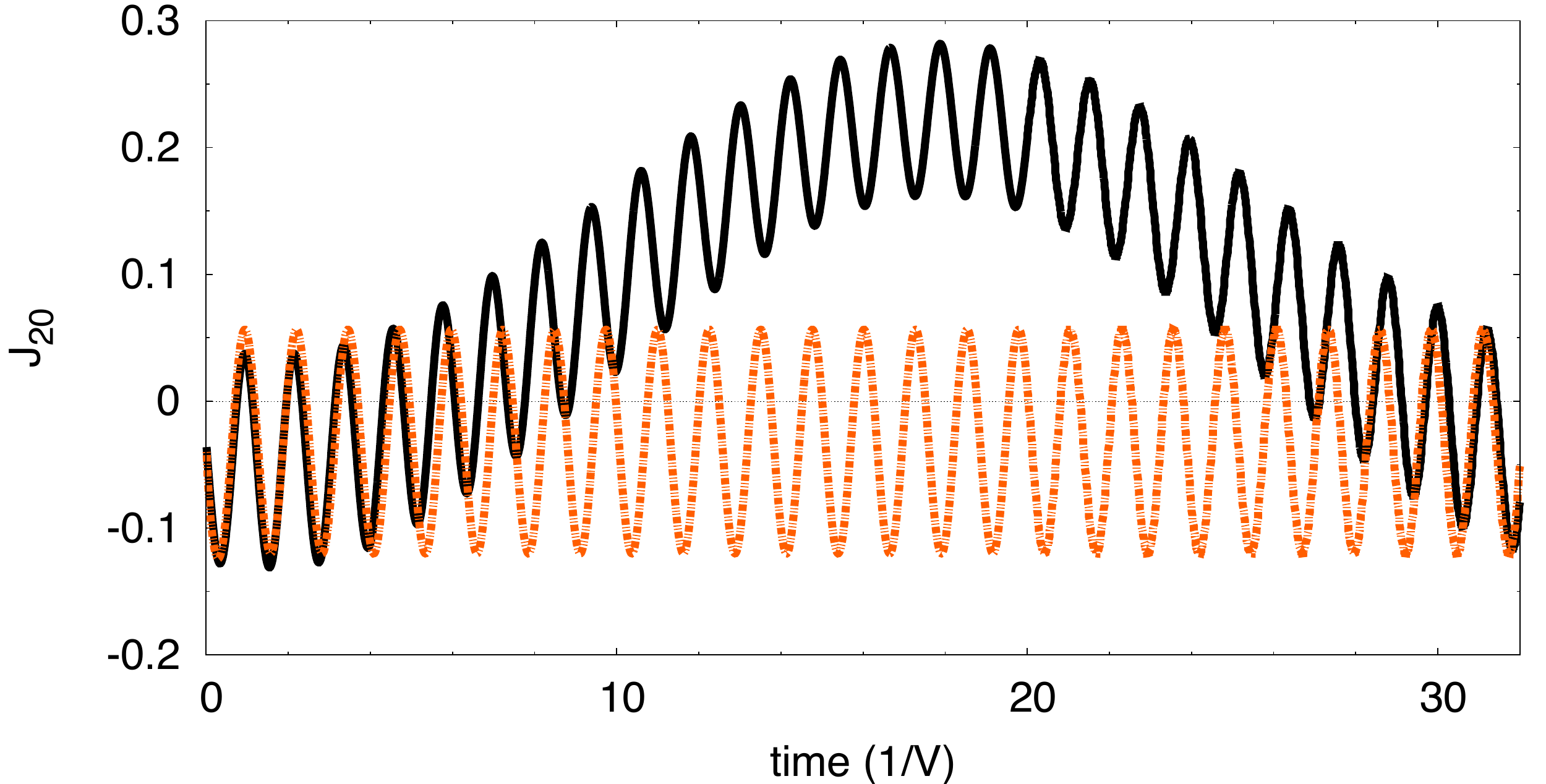}
\caption{\label{fig:J20DF-after} (Color online) Exact (black solid) and first-order (orange dashed) dynamics for $J_{20}$ after the quench.  Approximate result implied by Eq.~(\ref{eq:AA-after-approximate}).}
\end{figure}

Now let us carry the calculation to second order.  The second-order Hamiltonian, cf.~Eq.~(\ref{eq:E-S}), is 
\begin{equation*}
E_2(J) = \f{\partial H_1}{\partial J_{\mu}} \f{\partial S_1}{\partial \psi_{\mu0}} + \omega_{\mu0} \f{\partial S_2}{\partial \psi_{\mu0}} + \f{1}{2} \f{\partial \omega_{\mu0}}{\partial J_{\nu}} \f{\partial S_1}{\partial \psi_{\mu0}} \f{\partial S_1}{\partial \psi_{\nu0}},
\end{equation*}
where the last term vanishes because $\partial \omega_{\mu0}/\partial J_{\nu} = 0$.  Therefore, the differential equation for $S_2$ has a form similar to the one for $S_1$ in Eq.~(\ref{eq:S1-diffeq2}):
\begin{equation}
\omega_{10} \f{\partial S_2}{\partial \psi_{10}} + \omega_{20} \f{\partial S_2}{\partial \psi_{20}} = \langle H_2 \rangle - H_2, \label{eqn:S2}
\end{equation}
where we have averaged over $\psi_{10}$ and $\psi_{20}$ to define
\begin{align}
\langle H_2 \rangle &= \left< \sum_{\mu} \f{\partial H_1}{\partial J_{\mu}} \f{\partial S_1}{\partial \phi_{\mu0}} \right> \nonumber \\
&= \f{V^2}{2U} (-2 + 3Q_2).
\end{align}
The only nonzero Fourier components of the right-hand side of Eq.~(\ref{eqn:S2}) are
\begin{align*}
K_{0,1}^{(2)} = K_{0,-1}^{(2)} = K_{2,-1}^{(2)} = K_{-2,1}^{(2)} = \f{V^2}{4U} \sqrt{Q_2^2-Q_1^2}.
\end{align*}
A resonance has appeared:  the $(2,-1)$ and $(-2,1)$ terms are resonant because $\omega_{10} = -U$ and $\omega_{20} = -2U$.  As in the previous section, the next course of action is to perform a canonical transformation that isolates the resonant pair.  We shall not here proceed any further in this direction.  Figure~\ref{fig:J20DF-after} shows the first-order approximation for $J_{20}$ implied by Eq.~(\ref{eq:AA-after-approximate}).  
The first-order result is not as accurate as the first-order results from the previous section.  The approximation does not capture the drift in the guiding center of the oscillations.  This not a consequence of stopping at the first order \textit{per se} but rather of not taking into account the resonance.  One possible solution to this problem is to transform to resonance-adapted coordinates as we did in the previous section.  Alternatively, the Krylov-Bogoliubov averaging method can be used to derive approximations that take into account the drift in the center of the oscillations.  Further work is also needed to address the nontrivial dynamics precisely at the quench, where the conserved quantities jump suddenly to different values. 

\section{\label{sec:conclusions} Conclusions and Outlook}

Nonequilibrium dynamics is challenging because most of the many-body techniques we have were designed for equilibrium or steady states.  More and more experiments are probing nonlinear dynamical regimes that display unanticipated phenomena with no counterparts in equilibrium systems.  At the same time, there are many fundamental questions that remain to be addressed such as equilibration in strongly-interacting closed systems, the dynamics of quantum phase transitions and the influence of correlation and coherence in real-time dynamics.  There is still much to be understood about the physics of quantum many-body systems far from equilibrium. 

In this paper, a new approach to strongly-correlated nonequilibrium quantum dynamics has been presented.   It is based on the Hamiltonian structure of the BBGKY hierarchy for reduced density matrices.  Remarkably, the entire hierarchy of equations of motion can be expressed in the form of Hamilton equations for canonically conjugate variables, i.e.~generalized coordinates and momenta.  The resulting equations are just as intractable as the original ones, since the dimension of the resulting phase space is enormous.  However, expressing the equations in the form of Hamilton equations lets one bring to bear the well-developed approximation schemes of classical mechanics, for instance canonical perturbation theory and the Krylov-Bogoliubov averaging method, and in this way greatly reduce the dimension and complexity of the problem.  In Sec.~\ref{sec:quench}, canonical perturbation was applied to calculate the nontrivial quantum dynamics of a finite Hubbard chain which undergoes an interaction quench. 

Another way to effectively reduce the dimension of the problem is to close the equations by means of functional approximations.  
Here, I have put forward the conjecture that it is possible to close the BBGKY equations at any level of the hierarchy in the form of Hamilton equations for a complete set of canonically conjugate variables.  At the first level of the hierarchy, the complete set of variables $\{q_1^{\mu},p_{1\mu}\}$ contains the orbital degrees of freedom as well as the occupation numbers $n_k$ and their conjugate phases $\alpha_k$.  By accounting for the $\{\alpha_k,n_k\}$ dynamics, the approach goes beyond mean-field theory and brings us closer to describing strongly-correlated dynamics in the time domain.  A different functional theory, using essentially the same set of variables but leading to effective single-particle Schr\"odinger equations, has also been introduced \cite{giesbertz2010b}.  The most versatile approach to many-body nonequilibrium dynamics might be a hybrid approach, in which some of the degrees of freedom are described through the full hierarchy of BBGKY equations while others are eliminated through functional approximations.

The symplectic structure of the BBGKY hierarchy plays a key role in our formulation.  It is a prerequisite for Hamiltonian structure, and it guarantees the existence of a set of canonically conjugate variables.  Symplectic structure is also responsible for a new type of \textit{reduced} geometric phase, which is associated with cyclic evolutions of the reduced density matrices.  In contrast to Berry and Aharonov-Anandan \cite{berry1984,aharonov1987} phases, the reduced phases are observable even if the evolution of the full wave function is noncyclic.  The physical significance of these geometric phases remains to be explored.  Since they are sensitive to correlation and entanglement, they might lead to new insights into the dynamics of strongly correlated systems.  The reduced geometric phases can be expressed as action integrals of the form $\oint p_{\mu} dq^{\mu}$.  Action integrals play an important role in the transition from quantum mechanics to classical mechanics, and they often have a transparent physical meaning.  Another direction for future work is to investigate whether similar reduced geometric phases will appear for other types of reduced density matrices.  

The Hamiltonian structure of the equations is also important for another reason:  the great utility of canonical transformations in deriving systematic approximations and finding more convenient variables such as action-angle variables.  Even if action-angle variables do not exist, it is often possible to make a sequence of canonical transformations to variables that behave more and more like action-angle variables.  Apparently complicated dynamics can sometimes be described by slowly and regularly varying functions after such a transformation.  It is natural to suppose that the theoretical description of nonequilibrium dynamics will be facilitated by working in terms of the most slowly varying quantities.  This is especially important for the introduction of \textit{ab initio} functional approximations in the reduced density matrix equations of motion.  The main obstacle to applying TD DFT-like theories to real-time dynamics is the lack of knowledge about the memory dependence of the relevant functionals, such as the Kohn-Sham potential.  One can expect short-term memory dependence to be weaker when the functional $\mathcal{H}_n[Q_n^{\mu},P_{n\mu}]$, introduced in Eq.~(\ref{eq:conjecture}), is expressed in terms of optimal action-angle variables.  The Hamilton equations for the reduced variables $\{Q_n^{\mu},P_{n\mu}\}$ are in a form well-suited to \textit{ab initio} functional approximations.  Hamilton equations are also known to be a good starting point for setting up stable propagation algorithms that conserve energy and avoid secular terms, which can be expected to have important advantages in the simulation of slow transient and nonequilibrium processes such as relaxation and decoherence.  The mapping of quantum dynamics onto effective classical Hamilton equations might also facilitate the development of semiclassical approximations.  

The Hamiltonian formulation presented here for a system of bosons or fermions can be generalized to multi-component systems.  Classical analogs of quantum systems have been used for a long time in studying the semiclassical limit of nonadiabatic coupled electron-ion dynamics \cite{meyer1979,stock1997}.  Very recently, the classical Hamiltonian formulation of quantum degrees of freedom was used to study the structure of quantum-classical hybrid systems \cite{elze2012}.  Hybrid quantum-classical equations have also been derived by starting from the fully quantum equations and constraining the quantum fluctuations of a subsystem \cite{radonjic2012}.  Electron-ion dynamics is an example of a problem where approximations based on the separation of fast and slow degrees of freedom has a long and successful history, and it will be interesting to see whether the Hamiltonian formulation leads to further developments.

\begin{acknowledgments}
It is a pleasure to acknowledge helpful discussions with Alexander Abanov and Michele Fabrizio.
\end{acknowledgments}

\bibliography{bibliography2012a}

\begin{thebibliography}{88}
\expandafter\ifx\csname natexlab\endcsname\relax\def\natexlab#1{#1}\fi
\expandafter\ifx\csname bibnamefont\endcsname\relax
  \def\bibnamefont#1{#1}\fi
\expandafter\ifx\csname bibfnamefont\endcsname\relax
  \def\bibfnamefont#1{#1}\fi
\expandafter\ifx\csname citenamefont\endcsname\relax
  \def\citenamefont#1{#1}\fi
\expandafter\ifx\csname url\endcsname\relax
  \def\url#1{\texttt{#1}}\fi
\expandafter\ifx\csname urlprefix\endcsname\relax\def\urlprefix{URL }\fi
\providecommand{\bibinfo}[2]{#2}
\providecommand{\eprint}[2][]{\url{#2}}

\bibitem[{\citenamefont{Reider}(2004)}]{reider2004}
\bibinfo{author}{\bibfnamefont{G.~A.} \bibnamefont{Reider}},
  \bibinfo{journal}{J. Phys. D: Appl. Phys.} \textbf{\bibinfo{volume}{37}},
  \bibinfo{pages}{R37} (\bibinfo{year}{2004}).

\bibitem[{\citenamefont{Corkum and Krausz}(2007)}]{corkum2007}
\bibinfo{author}{\bibfnamefont{P.~B.} \bibnamefont{Corkum}} \bibnamefont{and}
  \bibinfo{author}{\bibfnamefont{F.}~\bibnamefont{Krausz}},
  \bibinfo{journal}{Nat. Phys.} \textbf{\bibinfo{volume}{3}},
  \bibinfo{pages}{381} (\bibinfo{year}{2007}).

\bibitem[{\citenamefont{Krausz and Ivanov}(2009)}]{krausz2009}
\bibinfo{author}{\bibfnamefont{F.}~\bibnamefont{Krausz}} \bibnamefont{and}
  \bibinfo{author}{\bibfnamefont{M.}~\bibnamefont{Ivanov}},
  \bibinfo{journal}{Rev. Mod. Phys.} \textbf{\bibinfo{volume}{81}},
  \bibinfo{pages}{163} (\bibinfo{year}{2009}).

\bibitem[{\citenamefont{Bloch}(2008)}]{bloch2008}
\bibinfo{author}{\bibfnamefont{I.}~\bibnamefont{Bloch}}, \bibinfo{journal}{Rev.
  Mod. Phys.} \textbf{\bibinfo{volume}{80}}, \bibinfo{pages}{885}
  (\bibinfo{year}{2008}).

\bibitem[{\citenamefont{Kirilyuk et~al.}(2010)\citenamefont{Kirilyuk, Kimel,
  and Rasing}}]{kirilyuk2010}
\bibinfo{author}{\bibfnamefont{A.}~\bibnamefont{Kirilyuk}},
  \bibinfo{author}{\bibfnamefont{A.~V.} \bibnamefont{Kimel}}, \bibnamefont{and}
  \bibinfo{author}{\bibfnamefont{T.}~\bibnamefont{Rasing}},
  \bibinfo{journal}{Rev. Mod. Phys.} \textbf{\bibinfo{volume}{82}},
  \bibinfo{pages}{2731} (\bibinfo{year}{2010}).

\bibitem[{\citenamefont{{I. Radu, et al.}}(2011)}]{radu2011}
\bibinfo{author}{\bibnamefont{{I. Radu, et al.}}}, \bibinfo{journal}{Nature}
  \textbf{\bibinfo{volume}{472}}, \bibinfo{pages}{205} (\bibinfo{year}{2011}).

\bibitem[{\citenamefont{Polkovnikov}(2005)}]{polkovnikov2005}
\bibinfo{author}{\bibfnamefont{A.}~\bibnamefont{Polkovnikov}},
  \bibinfo{journal}{Phys. Rev. B} \textbf{\bibinfo{volume}{72}},
  \bibinfo{pages}{161201(R)} (\bibinfo{year}{2005}).

\bibitem[{\citenamefont{Zurek et~al.}(2005)\citenamefont{Zurek, Dorner, and
  Zoller}}]{zurek2005}
\bibinfo{author}{\bibfnamefont{W.~H.} \bibnamefont{Zurek}},
  \bibinfo{author}{\bibfnamefont{U.}~\bibnamefont{Dorner}}, \bibnamefont{and}
  \bibinfo{author}{\bibfnamefont{P.}~\bibnamefont{Zoller}},
  \bibinfo{journal}{Phys. Rev. Lett.} \textbf{\bibinfo{volume}{95}},
  \bibinfo{pages}{105701} (\bibinfo{year}{2005}).

\bibitem[{\citenamefont{Cazalilla and Rigol}(2010)}]{cazalilla2010}
\bibinfo{author}{\bibfnamefont{M.~A.} \bibnamefont{Cazalilla}}
  \bibnamefont{and} \bibinfo{author}{\bibfnamefont{M.}~\bibnamefont{Rigol}},
  \bibinfo{journal}{New J. Phys.} \textbf{\bibinfo{volume}{12}},
  \bibinfo{pages}{055006} (\bibinfo{year}{2010}).

\bibitem[{\citenamefont{Biroli et~al.}(2010)\citenamefont{Biroli, Kollath, and
  Lauchli}}]{biroli2010}
\bibinfo{author}{\bibfnamefont{G.}~\bibnamefont{Biroli}},
  \bibinfo{author}{\bibfnamefont{C.}~\bibnamefont{Kollath}}, \bibnamefont{and}
  \bibinfo{author}{\bibfnamefont{A.~M.} \bibnamefont{Lauchli}},
  \bibinfo{journal}{Phys. Rev. Lett.} \textbf{\bibinfo{volume}{105}},
  \bibinfo{pages}{250401} (\bibinfo{year}{2010}).

\bibitem[{\citenamefont{Polkovnikov et~al.}(2011)\citenamefont{Polkovnikov,
  Sengupta, Silva, and Vengalattore}}]{polkovnikov2011}
\bibinfo{author}{\bibfnamefont{A.}~\bibnamefont{Polkovnikov}},
  \bibinfo{author}{\bibfnamefont{K.}~\bibnamefont{Sengupta}},
  \bibinfo{author}{\bibfnamefont{A.}~\bibnamefont{Silva}}, \bibnamefont{and}
  \bibinfo{author}{\bibfnamefont{M.}~\bibnamefont{Vengalattore}},
  \bibinfo{journal}{Rev. Mod. Phys.} \textbf{\bibinfo{volume}{83}},
  \bibinfo{pages}{863} (\bibinfo{year}{2011}).

\bibitem[{\citenamefont{Kadanoff and Baym}(1962)}]{kadanoff1962}
\bibinfo{author}{\bibfnamefont{L.~P.} \bibnamefont{Kadanoff}} \bibnamefont{and}
  \bibinfo{author}{\bibfnamefont{G.}~\bibnamefont{Baym}},
  \emph{\bibinfo{title}{Quantum Statistical Mechanics}}
  (\bibinfo{publisher}{Benjamin, New York}, \bibinfo{year}{1962}).

\bibitem[{\citenamefont{Keldysh}(1964)}]{keldysh1964}
\bibinfo{author}{\bibfnamefont{L.~V.} \bibnamefont{Keldysh}},
  \bibinfo{journal}{Zh. Eksp. Teor. Fiz.} \textbf{\bibinfo{volume}{47}},
  \bibinfo{pages}{1515} (\bibinfo{year}{1964}), \bibinfo{note}{[Sov. Phys. JETP
  \textbf{20}, 1018 (1965)]}.

\bibitem[{\citenamefont{Bogoliubov}(1946)}]{bogoliubov1946}
\bibinfo{author}{\bibfnamefont{N.~N.} \bibnamefont{Bogoliubov}},
  \emph{\bibinfo{title}{Problems of Dynamical Theory in Statistical Physics}}
  (\bibinfo{publisher}{Gostekhisdat}, \bibinfo{year}{1946}),
  \bibinfo{note}{(russ.)}.

\bibitem[{\citenamefont{de~Boer and Uhlenbeck}(1961)}]{bogoliubov1961a}
\bibinfo{editor}{\bibfnamefont{J.}~\bibnamefont{de~Boer}} \bibnamefont{and}
  \bibinfo{editor}{\bibfnamefont{G.~E.} \bibnamefont{Uhlenbeck}}, eds.,
  \emph{\bibinfo{title}{Studies in Statistical Mechanics}}
  (\bibinfo{publisher}{North-Holland, Amsterdam}, \bibinfo{year}{1961}),
  vol.~\bibinfo{volume}{1}.

\bibitem[{\citenamefont{Bonitz}(1998)}]{bonitz1998}
\bibinfo{author}{\bibfnamefont{M.}~\bibnamefont{Bonitz}},
  \emph{\bibinfo{title}{Quantum Kinetic Theory}} (\bibinfo{publisher}{B. G.
  Teubner, Stuttgart Leipzig}, \bibinfo{year}{1998}).

\bibitem[{\citenamefont{Runge and Gross}(1984)}]{runge1984}
\bibinfo{author}{\bibfnamefont{E.}~\bibnamefont{Runge}} \bibnamefont{and}
  \bibinfo{author}{\bibfnamefont{E.~K.~U.} \bibnamefont{Gross}},
  \bibinfo{journal}{Phys. Rev. Lett.} \textbf{\bibinfo{volume}{52}},
  \bibinfo{pages}{997} (\bibinfo{year}{1984}).

\bibitem[{\citenamefont{Freericks et~al.}(2006)\citenamefont{Freericks,
  Turkowski, and Zlatic}}]{freericks2006}
\bibinfo{author}{\bibfnamefont{J.~K.} \bibnamefont{Freericks}},
  \bibinfo{author}{\bibfnamefont{V.~M.} \bibnamefont{Turkowski}},
  \bibnamefont{and} \bibinfo{author}{\bibfnamefont{V.}~\bibnamefont{Zlatic}},
  \bibinfo{journal}{Phys. Rev. Lett.} \textbf{\bibinfo{volume}{97}},
  \bibinfo{pages}{266408} (\bibinfo{year}{2006}).

\bibitem[{\citenamefont{{M. Eckstein, et al.}}(2010)}]{eckstein2010}
\bibinfo{author}{\bibnamefont{{M. Eckstein, et al.}}}, \bibinfo{journal}{Eur.
  Phys. J. Special Topics} \textbf{\bibinfo{volume}{180}}, \bibinfo{pages}{217}
  (\bibinfo{year}{2010}).

\bibitem[{\citenamefont{Schir\'o and Fabrizio}(2010)}]{schiro2010}
\bibinfo{author}{\bibfnamefont{M.}~\bibnamefont{Schir\'o}} \bibnamefont{and}
  \bibinfo{author}{\bibfnamefont{M.}~\bibnamefont{Fabrizio}},
  \bibinfo{journal}{Phys. Rev. Lett.} \textbf{\bibinfo{volume}{105}},
  \bibinfo{pages}{076401} (\bibinfo{year}{2010}).

\bibitem[{\citenamefont{Schir\'o and Fabrizio}(2011)}]{schiro2011}
\bibinfo{author}{\bibfnamefont{M.}~\bibnamefont{Schir\'o}} \bibnamefont{and}
  \bibinfo{author}{\bibfnamefont{M.}~\bibnamefont{Fabrizio}},
  \bibinfo{journal}{Phys. Rev. B} \textbf{\bibinfo{volume}{83}},
  \bibinfo{pages}{165105} (\bibinfo{year}{2011}).

\bibitem[{\citenamefont{Schollwoeck}(2005)}]{schollwoeck2005}
\bibinfo{author}{\bibfnamefont{U.}~\bibnamefont{Schollwoeck}},
  \bibinfo{journal}{Rev. Mod. Phys.} \textbf{\bibinfo{volume}{77}},
  \bibinfo{pages}{259} (\bibinfo{year}{2005}).

\bibitem[{\citenamefont{{E. Gull, et al.}}(2011)}]{gull2011}
\bibinfo{author}{\bibnamefont{{E. Gull, et al.}}}, \bibinfo{journal}{Rev. Mod.
  Phys.} \textbf{\bibinfo{volume}{83}}, \bibinfo{pages}{349}
  (\bibinfo{year}{2011}).

\bibitem[{\citenamefont{Moeckel and Kehrein}(2010)}]{moeckel2010}
\bibinfo{author}{\bibfnamefont{M.}~\bibnamefont{Moeckel}} \bibnamefont{and}
  \bibinfo{author}{\bibfnamefont{S.}~\bibnamefont{Kehrein}},
  \bibinfo{journal}{New J. Phys.} \textbf{\bibinfo{volume}{12}},
  \bibinfo{pages}{055016} (\bibinfo{year}{2010}).

\bibitem[{\citenamefont{Bogoliubov and Mitropolsky}(1961)}]{bogoliubov1961b}
\bibinfo{author}{\bibfnamefont{N.~N.} \bibnamefont{Bogoliubov}}
  \bibnamefont{and} \bibinfo{author}{\bibfnamefont{Y.~A.}
  \bibnamefont{Mitropolsky}}, \emph{\bibinfo{title}{Asymptotic methods in the
  theory of non-linear oscillations}} (\bibinfo{publisher}{Gordan and Breach
  (NY)}, \bibinfo{year}{1961}).

\bibitem[{\citenamefont{Kevorkian and Cole}(1996)}]{kevorkian1996}
\bibinfo{author}{\bibfnamefont{J.}~\bibnamefont{Kevorkian}} \bibnamefont{and}
  \bibinfo{author}{\bibfnamefont{J.~D.} \bibnamefont{Cole}},
  \emph{\bibinfo{title}{Multiple scale and singular perturbation methods}}
  (\bibinfo{publisher}{Springer-Verlag (NY)}, \bibinfo{year}{1996}).

\bibitem[{\citenamefont{Verhulst}(2005)}]{verhulst2005}
\bibinfo{author}{\bibfnamefont{F.}~\bibnamefont{Verhulst}},
  \emph{\bibinfo{title}{Methods and applications of singular perturbations}}
  (\bibinfo{publisher}{Springer, New York}, \bibinfo{year}{2005}).

\bibitem[{\citenamefont{Smerzi et~al.}(1997)\citenamefont{Smerzi, Fantoni,
  Giovanazzi, and Shenoy}}]{smerzi1997}
\bibinfo{author}{\bibfnamefont{A.}~\bibnamefont{Smerzi}},
  \bibinfo{author}{\bibfnamefont{S.}~\bibnamefont{Fantoni}},
  \bibinfo{author}{\bibfnamefont{S.}~\bibnamefont{Giovanazzi}},
  \bibnamefont{and} \bibinfo{author}{\bibfnamefont{S.~R.}
  \bibnamefont{Shenoy}}, \bibinfo{journal}{Phys. Rev. Lett.}
  \textbf{\bibinfo{volume}{79}}, \bibinfo{pages}{4950} (\bibinfo{year}{1997}).

\bibitem[{\citenamefont{Zobay and Garraway}(2000)}]{zobay2000}
\bibinfo{author}{\bibfnamefont{O.}~\bibnamefont{Zobay}} \bibnamefont{and}
  \bibinfo{author}{\bibfnamefont{B.~M.} \bibnamefont{Garraway}},
  \bibinfo{journal}{Phys. Rev. A} \textbf{\bibinfo{volume}{61}},
  \bibinfo{pages}{033603} (\bibinfo{year}{2000}).

\bibitem[{\citenamefont{Graefe et~al.}(2006)\citenamefont{Graefe, Korsch, and
  Witthaut}}]{graefe2006}
\bibinfo{author}{\bibfnamefont{E.~M.} \bibnamefont{Graefe}},
  \bibinfo{author}{\bibfnamefont{H.~J.} \bibnamefont{Korsch}},
  \bibnamefont{and} \bibinfo{author}{\bibfnamefont{D.}~\bibnamefont{Witthaut}},
  \bibinfo{journal}{Phys. Rev. A} \textbf{\bibinfo{volume}{73}},
  \bibinfo{pages}{013617} (\bibinfo{year}{2006}).

\bibitem[{\citenamefont{Das et~al.}(2006)\citenamefont{Das, Sengupta, Sen, and
  Chakrabarti}}]{das2006}
\bibinfo{author}{\bibfnamefont{A.}~\bibnamefont{Das}},
  \bibinfo{author}{\bibfnamefont{K.}~\bibnamefont{Sengupta}},
  \bibinfo{author}{\bibfnamefont{D.}~\bibnamefont{Sen}}, \bibnamefont{and}
  \bibinfo{author}{\bibfnamefont{B.~K.} \bibnamefont{Chakrabarti}},
  \bibinfo{journal}{Phys. Rev. B} \textbf{\bibinfo{volume}{74}},
  \bibinfo{pages}{144423} (\bibinfo{year}{2006}).

\bibitem[{\citenamefont{Itin and Watanabe}(2007)}]{itin2007}
\bibinfo{author}{\bibfnamefont{A.~P.} \bibnamefont{Itin}} \bibnamefont{and}
  \bibinfo{author}{\bibfnamefont{S.}~\bibnamefont{Watanabe}},
  \bibinfo{journal}{Phys. Rev. E} \textbf{\bibinfo{volume}{76}},
  \bibinfo{pages}{026218} (\bibinfo{year}{2007}).

\bibitem[{\citenamefont{Altland et~al.}(2009)\citenamefont{Altland, Gurarie,
  Kriecherbauer, and Polkovnikov}}]{altland2009}
\bibinfo{author}{\bibfnamefont{A.}~\bibnamefont{Altland}},
  \bibinfo{author}{\bibfnamefont{V.}~\bibnamefont{Gurarie}},
  \bibinfo{author}{\bibfnamefont{T.}~\bibnamefont{Kriecherbauer}},
  \bibnamefont{and}
  \bibinfo{author}{\bibfnamefont{A.}~\bibnamefont{Polkovnikov}},
  \bibinfo{journal}{Phys. Rev. A} \textbf{\bibinfo{volume}{79}},
  \bibinfo{pages}{042703} (\bibinfo{year}{2009}).

\bibitem[{\citenamefont{Keeling et~al.}(2010)\citenamefont{Keeling, Bhaseen,
  and Simons}}]{keeling2010}
\bibinfo{author}{\bibfnamefont{J.}~\bibnamefont{Keeling}},
  \bibinfo{author}{\bibfnamefont{M.~J.} \bibnamefont{Bhaseen}},
  \bibnamefont{and} \bibinfo{author}{\bibfnamefont{B.~D.}
  \bibnamefont{Simons}}, \bibinfo{journal}{Phys. Rev. Lett.}
  \textbf{\bibinfo{volume}{105}}, \bibinfo{pages}{043001}
  (\bibinfo{year}{2010}).

\bibitem[{\citenamefont{Sciolla and Biroli}(2011)}]{sciolla2011}
\bibinfo{author}{\bibfnamefont{B.}~\bibnamefont{Sciolla}} \bibnamefont{and}
  \bibinfo{author}{\bibfnamefont{G.}~\bibnamefont{Biroli}},
  \bibinfo{journal}{J. Stat. Mech.: Theory and Experiment} p.
  \bibinfo{pages}{P11003} (\bibinfo{year}{2011}).

\bibitem[{\citenamefont{Provost and Vallee}(1980)}]{provost1980}
\bibinfo{author}{\bibfnamefont{J.~P.} \bibnamefont{Provost}} \bibnamefont{and}
  \bibinfo{author}{\bibfnamefont{G.}~\bibnamefont{Vallee}},
  \bibinfo{journal}{Commun. Math. Phys.} \textbf{\bibinfo{volume}{76}},
  \bibinfo{pages}{289} (\bibinfo{year}{1980}).

\bibitem[{\citenamefont{Berry}(1989)}]{berry1989}
\bibinfo{author}{\bibfnamefont{M.~V.} \bibnamefont{Berry}},
  \emph{\bibinfo{title}{The quantum phase, five years after}}
  (\bibinfo{year}{1989}), pp. \bibinfo{pages}{7--28}, \bibinfo{note}{in
  Ref.~\cite{shapere1989}}.

\bibitem[{\citenamefont{Carollo and Pachos}(2005)}]{carollo2005}
\bibinfo{author}{\bibfnamefont{A.~C.~M.} \bibnamefont{Carollo}}
  \bibnamefont{and} \bibinfo{author}{\bibfnamefont{J.~K.}
  \bibnamefont{Pachos}}, \bibinfo{journal}{Phys. Rev. Lett.}
  \textbf{\bibinfo{volume}{95}}, \bibinfo{pages}{157203}
  (\bibinfo{year}{2005}).

\bibitem[{\citenamefont{CamposVenuti and Zanardi}(2007)}]{campos-venuti2007}
\bibinfo{author}{\bibfnamefont{L.}~\bibnamefont{CamposVenuti}}
  \bibnamefont{and} \bibinfo{author}{\bibfnamefont{P.}~\bibnamefont{Zanardi}},
  \bibinfo{journal}{Phys. Rev. Lett.} \textbf{\bibinfo{volume}{99}},
  \bibinfo{pages}{095701} (\bibinfo{year}{2007}).

\bibitem[{\citenamefont{Tomka et~al.}(2012)\citenamefont{Tomka, Polkovnikov,
  and Gritsev}}]{tomka2012}
\bibinfo{author}{\bibfnamefont{M.}~\bibnamefont{Tomka}},
  \bibinfo{author}{\bibfnamefont{A.}~\bibnamefont{Polkovnikov}},
  \bibnamefont{and} \bibinfo{author}{\bibfnamefont{V.}~\bibnamefont{Gritsev}},
  \bibinfo{journal}{Phys. Rev. Lett.} \textbf{\bibinfo{volume}{108}},
  \bibinfo{pages}{080404} (\bibinfo{year}{2012}).

\bibitem[{\citenamefont{Mead and Truhlar}(1979)}]{mead1979}
\bibinfo{author}{\bibfnamefont{C.~A.} \bibnamefont{Mead}} \bibnamefont{and}
  \bibinfo{author}{\bibfnamefont{D.~G.} \bibnamefont{Truhlar}},
  \bibinfo{journal}{J. Phys. Chem.} \textbf{\bibinfo{volume}{70}},
  \bibinfo{pages}{2284} (\bibinfo{year}{1979}).

\bibitem[{\citenamefont{Jackiw}(1988)}]{jackiw1988}
\bibinfo{author}{\bibfnamefont{R.}~\bibnamefont{Jackiw}},
  \bibinfo{journal}{Commun. At. Molec. Phys.} \textbf{\bibinfo{volume}{21}},
  \bibinfo{pages}{71} (\bibinfo{year}{1988}).

\bibitem[{\citenamefont{Goldhaber}(2005)}]{goldhaber2005}
\bibinfo{author}{\bibfnamefont{A.~S.} \bibnamefont{Goldhaber}},
  \bibinfo{journal}{Phys. Rev. A} \textbf{\bibinfo{volume}{71}},
  \bibinfo{pages}{062102} (\bibinfo{year}{2005}).

\bibitem[{\citenamefont{Abedi et~al.}(2010)\citenamefont{Abedi, Maitra, and
  Gross}}]{abedi2010}
\bibinfo{author}{\bibfnamefont{A.}~\bibnamefont{Abedi}},
  \bibinfo{author}{\bibfnamefont{N.~T.} \bibnamefont{Maitra}},
  \bibnamefont{and} \bibinfo{author}{\bibfnamefont{E.~K.~U.}
  \bibnamefont{Gross}}, \bibinfo{journal}{Phys. Rev. Lett.}
  \textbf{\bibinfo{volume}{105}}, \bibinfo{pages}{123002}
  (\bibinfo{year}{2010}).

\bibitem[{\citenamefont{Requist and Pankratov}(2010)}]{requist2010}
\bibinfo{author}{\bibfnamefont{R.}~\bibnamefont{Requist}} \bibnamefont{and}
  \bibinfo{author}{\bibfnamefont{O.}~\bibnamefont{Pankratov}},
  \bibinfo{journal}{Phys. Rev. A} \textbf{\bibinfo{volume}{81}},
  \bibinfo{pages}{042519} (\bibinfo{year}{2010}).

\bibitem[{\citenamefont{Gasparinetti et~al.}(2011)\citenamefont{Gasparinetti,
  Solinas, and Pekola}}]{gasparinetti2011}
\bibinfo{author}{\bibfnamefont{S.}~\bibnamefont{Gasparinetti}},
  \bibinfo{author}{\bibfnamefont{P.}~\bibnamefont{Solinas}}, \bibnamefont{and}
  \bibinfo{author}{\bibfnamefont{J.~P.} \bibnamefont{Pekola}},
  \bibinfo{journal}{Phys. Rev. Lett.} \textbf{\bibinfo{volume}{107}},
  \bibinfo{pages}{207002} (\bibinfo{year}{2011}).

\bibitem[{\citenamefont{Requist and Pankratov}(2011)}]{requist2011}
\bibinfo{author}{\bibfnamefont{R.}~\bibnamefont{Requist}} \bibnamefont{and}
  \bibinfo{author}{\bibfnamefont{O.}~\bibnamefont{Pankratov}},
  \bibinfo{journal}{Phys. Rev. A} \textbf{\bibinfo{volume}{83}},
  \bibinfo{pages}{052510} (\bibinfo{year}{2011}).

\bibitem[{\citenamefont{L\"owdin}(1955)}]{loewdin1955}
\bibinfo{author}{\bibfnamefont{P.~O.} \bibnamefont{L\"owdin}},
  \bibinfo{journal}{Phys. Rev.} \textbf{\bibinfo{volume}{97}},
  \bibinfo{pages}{1474} (\bibinfo{year}{1955}).

\bibitem[{\citenamefont{Jose and Saletan}(1998)}]{jose1998}
\bibinfo{author}{\bibfnamefont{J.~V.} \bibnamefont{Jose}} \bibnamefont{and}
  \bibinfo{author}{\bibfnamefont{E.~J.} \bibnamefont{Saletan}},
  \emph{\bibinfo{title}{Classical dynamics}} (\bibinfo{publisher}{Cambridge},
  \bibinfo{year}{1998}).

\bibitem[{\citenamefont{Dirac}(1927)}]{dirac1927}
\bibinfo{author}{\bibfnamefont{P.~A.~M.} \bibnamefont{Dirac}},
  \bibinfo{journal}{Proc. Roy. Soc. Lond. A} \textbf{\bibinfo{volume}{114}},
  \bibinfo{pages}{243} (\bibinfo{year}{1927}).

\bibitem[{\citenamefont{Kibble}(1979)}]{kibble1979}
\bibinfo{author}{\bibfnamefont{T.~W.~B.} \bibnamefont{Kibble}},
  \bibinfo{journal}{Commun. Math. Phys.} \textbf{\bibinfo{volume}{65}},
  \bibinfo{pages}{189} (\bibinfo{year}{1979}).

\bibitem[{\citenamefont{Rowe et~al.}(1980)\citenamefont{Rowe, Ryman, and
  Rosensteel}}]{rowe1980}
\bibinfo{author}{\bibfnamefont{D.~J.} \bibnamefont{Rowe}},
  \bibinfo{author}{\bibfnamefont{A.}~\bibnamefont{Ryman}}, \bibnamefont{and}
  \bibinfo{author}{\bibfnamefont{G.}~\bibnamefont{Rosensteel}},
  \bibinfo{journal}{Phys. Rev. A} \textbf{\bibinfo{volume}{22}},
  \bibinfo{pages}{2362} (\bibinfo{year}{1980}).

\bibitem[{\citenamefont{Heslot}(1985)}]{heslot1985}
\bibinfo{author}{\bibfnamefont{A.}~\bibnamefont{Heslot}},
  \bibinfo{journal}{Phys. Rev. D} \textbf{\bibinfo{volume}{31}},
  \bibinfo{pages}{1341} (\bibinfo{year}{1985}).

\bibitem[{\citenamefont{Anandan}(1991)}]{anandan1991}
\bibinfo{author}{\bibfnamefont{J.}~\bibnamefont{Anandan}},
  \bibinfo{journal}{Found. Phys.} \textbf{\bibinfo{volume}{21}},
  \bibinfo{pages}{1265} (\bibinfo{year}{1991}).

\bibitem[{\citenamefont{Chruscinski}(1991)}]{chruscinski1991}
\bibinfo{author}{\bibfnamefont{D.}~\bibnamefont{Chruscinski}},
  \bibinfo{journal}{Rep. Math. Phys.} \textbf{\bibinfo{volume}{29}},
  \bibinfo{pages}{95} (\bibinfo{year}{1991}).

\bibitem[{\citenamefont{Marsden et~al.}(1984)\citenamefont{Marsden, Morrison,
  and Weinstein}}]{marsden1984}
\bibinfo{author}{\bibfnamefont{J.~E.} \bibnamefont{Marsden}},
  \bibinfo{author}{\bibfnamefont{F.~J.} \bibnamefont{Morrison}},
  \bibnamefont{and}
  \bibinfo{author}{\bibfnamefont{A.}~\bibnamefont{Weinstein}},
  \bibinfo{journal}{Cont. Math. AMS} \textbf{\bibinfo{volume}{28}},
  \bibinfo{pages}{115} (\bibinfo{year}{1984}).

\bibitem[{\citenamefont{Guillemin and Sternberg}(1980)}]{guillemin1980}
\bibinfo{author}{\bibfnamefont{V.}~\bibnamefont{Guillemin}} \bibnamefont{and}
  \bibinfo{author}{\bibfnamefont{S.}~\bibnamefont{Sternberg}},
  \bibinfo{journal}{Ann. of Phys.} \textbf{\bibinfo{volume}{127}},
  \bibinfo{pages}{220} (\bibinfo{year}{1980}).

\bibitem[{\citenamefont{Carruthers and Nieto}(1968)}]{carruthers1968}
\bibinfo{author}{\bibfnamefont{P.}~\bibnamefont{Carruthers}} \bibnamefont{and}
  \bibinfo{author}{\bibfnamefont{M.~M.} \bibnamefont{Nieto}},
  \bibinfo{journal}{Rev. Mod. Phys.} \textbf{\bibinfo{volume}{40}},
  \bibinfo{pages}{411} (\bibinfo{year}{1968}).

\bibitem[{\citenamefont{Giesbertz
  et~al.}(2010{\natexlab{a}})\citenamefont{Giesbertz, Gritsenko, and
  Baerends}}]{giesbertz2010b}
\bibinfo{author}{\bibfnamefont{K.~J.~H.} \bibnamefont{Giesbertz}},
  \bibinfo{author}{\bibfnamefont{O.~V.} \bibnamefont{Gritsenko}},
  \bibnamefont{and} \bibinfo{author}{\bibfnamefont{E.~J.}
  \bibnamefont{Baerends}}, \bibinfo{journal}{Phys. Rev. Lett.}
  \textbf{\bibinfo{volume}{105}}, \bibinfo{pages}{013002}
  (\bibinfo{year}{2010}{\natexlab{a}}).

\bibitem[{\citenamefont{Coleman}(1963)}]{coleman1963}
\bibinfo{author}{\bibfnamefont{A.~J.} \bibnamefont{Coleman}},
  \bibinfo{journal}{Rev. Mod. Phys.} \textbf{\bibinfo{volume}{35}},
  \bibinfo{pages}{668} (\bibinfo{year}{1963}).

\bibitem[{\citenamefont{Klyachko}(2004)}]{klyachko2004}
\bibinfo{author}{\bibfnamefont{A.}~\bibnamefont{Klyachko}},
  \bibinfo{howpublished}{arxiv:0409113} (\bibinfo{year}{2004}).

\bibitem[{\citenamefont{Klyachko}(2006)}]{klyachko2006}
\bibinfo{author}{\bibfnamefont{A.~A.} \bibnamefont{Klyachko}},
  \bibinfo{journal}{J. Phys.: Conf. Series} \textbf{\bibinfo{volume}{36}},
  \bibinfo{pages}{72} (\bibinfo{year}{2006}).

\bibitem[{\citenamefont{Christandl et~al.}(2012)\citenamefont{Christandl,
  Doran, Kousidis, and Walter}}]{christandl2012}
\bibinfo{author}{\bibfnamefont{M.}~\bibnamefont{Christandl}},
  \bibinfo{author}{\bibfnamefont{B.}~\bibnamefont{Doran}},
  \bibinfo{author}{\bibfnamefont{S.}~\bibnamefont{Kousidis}}, \bibnamefont{and}
  \bibinfo{author}{\bibfnamefont{M.}~\bibnamefont{Walter}},
  \bibinfo{howpublished}{arxiv:1204.0741v1} (\bibinfo{year}{2012}).

\bibitem[{\citenamefont{Ashtekar and Schilling}(1998)}]{ashtekar1998}
\bibinfo{author}{\bibfnamefont{A.}~\bibnamefont{Ashtekar}} \bibnamefont{and}
  \bibinfo{author}{\bibfnamefont{T.~A.} \bibnamefont{Schilling}},
  \emph{\bibinfo{title}{On Einstein's path}}
  (\bibinfo{publisher}{Springer-Verlag, Berlin}, \bibinfo{year}{1998}),
  chap.~\bibinfo{chapter}{3}.

\bibitem[{\citenamefont{Ashtekar and Schilling}(1997)}]{ashtekar1997}
\bibinfo{author}{\bibfnamefont{A.}~\bibnamefont{Ashtekar}} \bibnamefont{and}
  \bibinfo{author}{\bibfnamefont{T.~A.} \bibnamefont{Schilling}},
  \bibinfo{howpublished}{arxiv:gr-qc/9706069} (\bibinfo{year}{1997}).

\bibitem[{\citenamefont{Berry}(1984)}]{berry1984}
\bibinfo{author}{\bibfnamefont{M.~V.} \bibnamefont{Berry}},
  \bibinfo{journal}{Proc. Roy. Soc. Lond. A} \textbf{\bibinfo{volume}{392}},
  \bibinfo{pages}{45} (\bibinfo{year}{1984}).

\bibitem[{\citenamefont{Aharonov and Anandan}(1987)}]{aharonov1987}
\bibinfo{author}{\bibfnamefont{Y.}~\bibnamefont{Aharonov}} \bibnamefont{and}
  \bibinfo{author}{\bibfnamefont{J.}~\bibnamefont{Anandan}},
  \bibinfo{journal}{Phys. Rev. Lett.} \textbf{\bibinfo{volume}{58}},
  \bibinfo{pages}{1593} (\bibinfo{year}{1987}).

\bibitem[{\citenamefont{Uhlmann}(1986)}]{uhlmann1986}
\bibinfo{author}{\bibfnamefont{A.}~\bibnamefont{Uhlmann}},
  \bibinfo{journal}{Rep. Math. Phys.} \textbf{\bibinfo{volume}{24}},
  \bibinfo{pages}{229} (\bibinfo{year}{1986}).

\bibitem[{\citenamefont{{E. Sj\"oqvist, et al.}}(2000)}]{sjoqvist2000a}
\bibinfo{author}{\bibnamefont{{E. Sj\"oqvist, et al.}}},
  \bibinfo{journal}{Phys. Rev. Lett.} \textbf{\bibinfo{volume}{85}},
  \bibinfo{pages}{2845} (\bibinfo{year}{2000}).

\bibitem[{\citenamefont{Sj\"oqvist}(2000)}]{sjoqvist2000b}
\bibinfo{author}{\bibfnamefont{E.}~\bibnamefont{Sj\"oqvist}},
  \bibinfo{journal}{Phys. Rev. A} \textbf{\bibinfo{volume}{62}},
  \bibinfo{pages}{022109} (\bibinfo{year}{2000}).

\bibitem[{\citenamefont{Samuel and Bhandari}(1988)}]{samuel1988}
\bibinfo{author}{\bibfnamefont{J.}~\bibnamefont{Samuel}} \bibnamefont{and}
  \bibinfo{author}{\bibfnamefont{R.}~\bibnamefont{Bhandari}},
  \bibinfo{journal}{Phys. Rev. Lett.} \textbf{\bibinfo{volume}{60}},
  \bibinfo{pages}{2339} (\bibinfo{year}{1988}).

\bibitem[{\citenamefont{Chaturvedi et~al.}(2004)\citenamefont{Chaturvedi,
  Ercolessi, Marmo, Morandi, Mukunda, and Simon}}]{chaturvedi2004}
\bibinfo{author}{\bibfnamefont{S.}~\bibnamefont{Chaturvedi}},
  \bibinfo{author}{\bibfnamefont{E.}~\bibnamefont{Ercolessi}},
  \bibinfo{author}{\bibfnamefont{G.}~\bibnamefont{Marmo}},
  \bibinfo{author}{\bibfnamefont{G.}~\bibnamefont{Morandi}},
  \bibinfo{author}{\bibfnamefont{N.}~\bibnamefont{Mukunda}}, \bibnamefont{and}
  \bibinfo{author}{\bibfnamefont{R.}~\bibnamefont{Simon}},
  \bibinfo{journal}{Eur. Phys. J. C} \textbf{\bibinfo{volume}{35}},
  \bibinfo{pages}{413} (\bibinfo{year}{2004}).

\bibitem[{\citenamefont{Itin and {T\"orm\"a}}(2009)}]{itin2009}
\bibinfo{author}{\bibfnamefont{A.~P.} \bibnamefont{Itin}} \bibnamefont{and}
  \bibinfo{author}{\bibfnamefont{P.}~\bibnamefont{{T\"orm\"a}}},
  \bibinfo{journal}{Phys. Rev. A} \textbf{\bibinfo{volume}{79}},
  \bibinfo{pages}{055602} (\bibinfo{year}{2009}).

\bibitem[{\citenamefont{Itin and Schmelcher}(2011)}]{itin2011}
\bibinfo{author}{\bibfnamefont{A.~P.} \bibnamefont{Itin}} \bibnamefont{and}
  \bibinfo{author}{\bibfnamefont{P.}~\bibnamefont{Schmelcher}},
  \bibinfo{journal}{Phys. Rev. A} \textbf{\bibinfo{volume}{84}},
  \bibinfo{pages}{063609} (\bibinfo{year}{2011}).

\bibitem[{\citenamefont{Cary}(1981)}]{cary1981}
\bibinfo{author}{\bibfnamefont{J.~R.} \bibnamefont{Cary}},
  \bibinfo{journal}{Phys. Rep.} \textbf{\bibinfo{volume}{79}},
  \bibinfo{pages}{129} (\bibinfo{year}{1981}).

\bibitem[{\citenamefont{{Van Kampen}}(1985)}]{vankampen1985}
\bibinfo{author}{\bibfnamefont{N.~G.} \bibnamefont{{Van Kampen}}},
  \bibinfo{journal}{Physics Reports} \textbf{\bibinfo{volume}{124}},
  \bibinfo{pages}{69} (\bibinfo{year}{1985}).

\bibitem[{\citenamefont{Nakajima}(1958)}]{nakajima1958}
\bibinfo{author}{\bibfnamefont{S.}~\bibnamefont{Nakajima}},
  \bibinfo{journal}{Prog. Theor. Phys.} \textbf{\bibinfo{volume}{20}},
  \bibinfo{pages}{948} (\bibinfo{year}{1958}).

\bibitem[{\citenamefont{Zwanzig}(1960)}]{zwanzig1960}
\bibinfo{author}{\bibfnamefont{R.}~\bibnamefont{Zwanzig}}, \bibinfo{journal}{J.
  Chem. Phys.} \textbf{\bibinfo{volume}{33}}, \bibinfo{pages}{1338}
  (\bibinfo{year}{1960}).

\bibitem[{\citenamefont{Giesbertz et~al.}(2012)\citenamefont{Giesbertz,
  Gritsenko, and Baerends}}]{giesbertz2012}
\bibinfo{author}{\bibfnamefont{K.~J.~H.} \bibnamefont{Giesbertz}},
  \bibinfo{author}{\bibfnamefont{O.~V.} \bibnamefont{Gritsenko}},
  \bibnamefont{and} \bibinfo{author}{\bibfnamefont{E.~J.}
  \bibnamefont{Baerends}}, \bibinfo{journal}{J. Chem. Phys.}
  \textbf{\bibinfo{volume}{136}}, \bibinfo{pages}{094104}
  (\bibinfo{year}{2012}).

\bibitem[{\citenamefont{Giesbertz
  et~al.}(2010{\natexlab{b}})\citenamefont{Giesbertz, Gritsenko, and
  Baerends}}]{giesbertz2010c}
\bibinfo{author}{\bibfnamefont{K.~J.~H.} \bibnamefont{Giesbertz}},
  \bibinfo{author}{\bibfnamefont{O.~V.} \bibnamefont{Gritsenko}},
  \bibnamefont{and} \bibinfo{author}{\bibfnamefont{E.~J.}
  \bibnamefont{Baerends}}, \bibinfo{journal}{J. Chem. Phys.}
  \textbf{\bibinfo{volume}{133}}, \bibinfo{pages}{174119}
  (\bibinfo{year}{2010}{\natexlab{b}}).

\bibitem[{\citenamefont{Hannay}(1985)}]{hannay1985}
\bibinfo{author}{\bibfnamefont{J.~H.} \bibnamefont{Hannay}},
  \bibinfo{journal}{J. Phys. A: Math. Gen.} \textbf{\bibinfo{volume}{18}},
  \bibinfo{pages}{221} (\bibinfo{year}{1985}).

\bibitem[{\citenamefont{Gross and Kohn}(1985)}]{gross1985}
\bibinfo{author}{\bibfnamefont{E.~K.~U.} \bibnamefont{Gross}} \bibnamefont{and}
  \bibinfo{author}{\bibfnamefont{W.}~\bibnamefont{Kohn}},
  \bibinfo{journal}{Phys. Rev. Lett.} \textbf{\bibinfo{volume}{55}},
  \bibinfo{pages}{2850} (\bibinfo{year}{1985}).

\bibitem[{\citenamefont{Vignale and Kohn}(1996)}]{vignale1996}
\bibinfo{author}{\bibfnamefont{G.}~\bibnamefont{Vignale}} \bibnamefont{and}
  \bibinfo{author}{\bibfnamefont{W.}~\bibnamefont{Kohn}},
  \bibinfo{journal}{Phys. Rev. Lett.} \textbf{\bibinfo{volume}{77}},
  \bibinfo{pages}{2037} (\bibinfo{year}{1996}).

\bibitem[{\citenamefont{Meyer and Miller}(1979)}]{meyer1979}
\bibinfo{author}{\bibfnamefont{H.-D.} \bibnamefont{Meyer}} \bibnamefont{and}
  \bibinfo{author}{\bibfnamefont{W.~H.} \bibnamefont{Miller}},
  \bibinfo{journal}{J. Chem. Phys} \textbf{\bibinfo{volume}{70}},
  \bibinfo{pages}{3214} (\bibinfo{year}{1979}).

\bibitem[{\citenamefont{Stock and Thoss}(1997)}]{stock1997}
\bibinfo{author}{\bibfnamefont{G.}~\bibnamefont{Stock}} \bibnamefont{and}
  \bibinfo{author}{\bibfnamefont{M.}~\bibnamefont{Thoss}},
  \bibinfo{journal}{Phys. Rev. Lett.} \textbf{\bibinfo{volume}{78}},
  \bibinfo{pages}{578} (\bibinfo{year}{1997}).

\bibitem[{\citenamefont{Elze}(2012)}]{elze2012}
\bibinfo{author}{\bibfnamefont{H.-T.} \bibnamefont{Elze}},
  \bibinfo{journal}{Phys. Rev. A} \textbf{\bibinfo{volume}{85}},
  \bibinfo{pages}{052109} (\bibinfo{year}{2012}).

\bibitem[{\citenamefont{Radonji\'c et~al.}(2012)\citenamefont{Radonji\'c,
  Prvanovi\'c, and Buri\'c}}]{radonjic2012}
\bibinfo{author}{\bibfnamefont{M.}~\bibnamefont{Radonji\'c}},
  \bibinfo{author}{\bibfnamefont{S.}~\bibnamefont{Prvanovi\'c}},
  \bibnamefont{and} \bibinfo{author}{\bibfnamefont{N.}~\bibnamefont{Buri\'c}},
  \bibinfo{journal}{Phys. Rev. A} \textbf{\bibinfo{volume}{85}},
  \bibinfo{pages}{064101} (\bibinfo{year}{2012}).

\bibitem[{\citenamefont{Shapere and Wilczek}(1989)}]{shapere1989}
\bibinfo{editor}{\bibfnamefont{A.}~\bibnamefont{Shapere}} \bibnamefont{and}
  \bibinfo{editor}{\bibfnamefont{F.}~\bibnamefont{Wilczek}}, eds.,
  \emph{\bibinfo{title}{Geometric phases in physics}}
  (\bibinfo{publisher}{World Scientific, Singapore}, \bibinfo{year}{1989}).

\end{thebibliography}

\end{document}